\documentclass[prb,showpacs,floatfix,superscriptaddress,twocolumn,aps,10,groupedaddress]{revtex4-2}

\usepackage{times}
\usepackage{multirow}
\usepackage[dvipsnames]{xcolor}
\usepackage{amsmath}
\usepackage{amsthm, mathrsfs}
\usepackage{amssymb}
\usepackage{amsbsy}
\usepackage{enumitem}
\usepackage{bm,comment}
\usepackage{wasysym}
\usepackage[english]{babel}
\usepackage[T1]{fontenc}
\usepackage[utf8]{inputenc} 
\usepackage[colorlinks,bookmarks=false,citecolor=blue,linkcolor=red,urlcolor=blue]{hyperref}
\usepackage{pstricks}
\usepackage{rotating}			       
\usepackage{tabularx,hhline}	
\usepackage[normalem]{ulem}
\usepackage{subcaption}
\usepackage{graphicx}

\newcolumntype{P}[1]{>{\centering\arraybackslash}p{#1}}

\newcommand\redsout{\bgroup\markoverwith{\textcolor{red}{\rule[0.5ex]{2pt}{0.4pt}}}\ULon}
\newcommand\bluesout{\bgroup\markoverwith{\textcolor{blue}{\rule[0.5ex]{2pt}{0.4pt}}}\ULon}

\newcommand{\SPhide}[1]{}

\begin{document}
\title{Diodes and capacitors for the transport of monopoles in fragmented spin ice}
\author{Anoop Raj}
\affiliation{Department of Physics, Indian Institute of Technology Bombay, Mumbai, MH 400076, India}
\email{sumiran.pujari@iitb.ac.in}

\author{Sumiran Pujari}
\affiliation{Department of Physics, Indian Institute of Technology Bombay, Mumbai, MH 400076, India}
\email{sumiran.pujari@iitb.ac.in}

\author{Ludovic D.C. Jaubert}
\affiliation{CNRS, Universit\'e de Bordeaux, LOMA, UMR 5798, 33400 Talence, France} 
\email{ludovic.jaubert@u-bordeaux.fr}
\date{\today}
\begin{abstract}
Spin-ice materials are famous for their quasi-particle excitations that behave like magnetic monopoles. Magnetricity is the concept that these monopoles can conduct an AC magnetic current, in analogy with conduction electrons. While monopole dynamics has been intensively studied and is reasonably well understood, very little has been done to design devices in order to control magnetricity. Here we develop a theoretical proof of concept for the design of diodes and capacitors for the transport of monopoles. We use the property of systems with magnetic fragmentation, where spin-ice physics co-exists with long-range antiferromagnetic order. The key point is that magnetic order allows for the existence of domain walls. Under certain conditions of preparation, this domain wall is equivalent to an asymmetric filter for monopoles. In a given direction, positive charges can go through while negative ones are repelled; the opposite applies in the opposite direction. This asymmetry effectively functions like a diode for monopole current. Successive domain walls separate positive from negative charges with a vacuum of charge in between, producing a capacitor for monopoles. Once the capacitor is charged, it can in principle be used as a battery for monopoles. All microscopic mechanisms are explained and our proof of concept is validated by simulations of more than a million spins. Application to experiments are discussed for rare-earth pyrochlore oxides and artificial spin ice. Finally, we discuss in general terms how a domain wall in fragmented spin ice can also be seen as an emergent boundary separating two mirror “worlds” separated by time-reversal symmetry. Beyond spin ice, our work opens a promising direction of investigation for the dynamics of emergent quasi-particles crossing domain walls in chiral and nematic spin liquids, which also possess a broken symmetry.
\end{abstract}
\maketitle

\section{Introduction}
\label{sec:intro}

The existence of magnetic monopoles as one of the elementary or fundamental particles has eluded observation till now.
However, magnetic monopoles are by now well established quasi-particles in spin-ice materials such as Dy$_2$Ti$_2$O$_7$ and Ho$_2$Ti$_2$O$_7$\cite{castelnovo08a,jaubert09a,morris09a,giblin11a,Kaiser18a,spinicebook}.
Emerging from the many-body behaviour of large rare-earth magnetic moments ($\sim 10 \mu_B$) in these materials, ultra-sensitive SQUID measurements have listened to the noise in their dynamics~\cite{Dusad19a,Samarakoon22a,Hallen22a}. 
Their low-temperature properties are also well understood within the Debye-H\"uckel theory of an electrolyte~\cite{castelnovo11a,giblin11a,jaubert11c,Kaiser18a}. 
This is why, as the closest magnetic counterpart of electrons, magnetic monopoles have raised the question of their transport; a concept coined ``magnetricity'' by Steve Bramwell and co-workers~\cite{bramwell09a,giblin11a}, that has been further extended from low-temperature solid-state crystals to room-temperature artificial lattices~\cite{mengotti11a,nisoli13a,kapaklis14a,perrin16a,farhan19a,spinicebook}. 
Monopoles differ from conduction electrons though, by the presence of Dirac strings between them~\cite{castelnovo08a,jaubert09a,morris09a}.  
While disordered and fluctuating in zero field, an external magnetic field confers a tension to the strings responsible for the magnetisation of the samples. This magnetisation prevents a DC current of monopoles, but allows for an AC one~\cite{jaubert09a,giblin11a}.

Once there is a current, the question then becomes how to control it. 
From this point of view, magnetricity is in its early days. 
Basic devices such as resistors, capacitors or diodes have not yet been designed, nor is it known if they are even theoretically possible. 
It is the goal of this paper to embrace this atypical route for spintronics, and to develop a theoretical proof of concept for some of these devices.

This route comes, however, with a delicate challenge. 
A system -- any system -- has to remain magnetically neutral, i.e. there must always be as many positive as negative magnetic monopoles (including boundary effects~\cite{jaubert09a,Miao20a}). 
In the language of electronics, how to $p$-dope ($n$-dope) a spin-ice crystal with positive (negative) magnetic carriers remains a difficult problem. 
Without the analogue of extrinsic semiconductors, can we still design active devices such as a diode ?
This is why we will not try to reproduce the microscopic physics of a $p$-$n$ junction here. Instead we shall take advantage of an alternative facet of spin ice, namely magnetic fragmentation~\cite{brooks14a}. 
As opposed to traditional spin ice, magnetic fragmentation represents the co-existence, in the same phase, of spin-ice physics with long-range magnetic order. 
Since there is magnetic order, domain walls are now possible in an extensively degenerate spin-ice manifold. 
But what happens when a monopole hits this domain wall ? How is its dynamics affected, and is it possible to control the transport of these magnetic carriers via domain walls ? These are some of the questions we will answer in this paper.

\begin{figure*}[t]
\centering
\begin{subfigure}[b]{0.22\textwidth}
	\includegraphics[width=\linewidth]{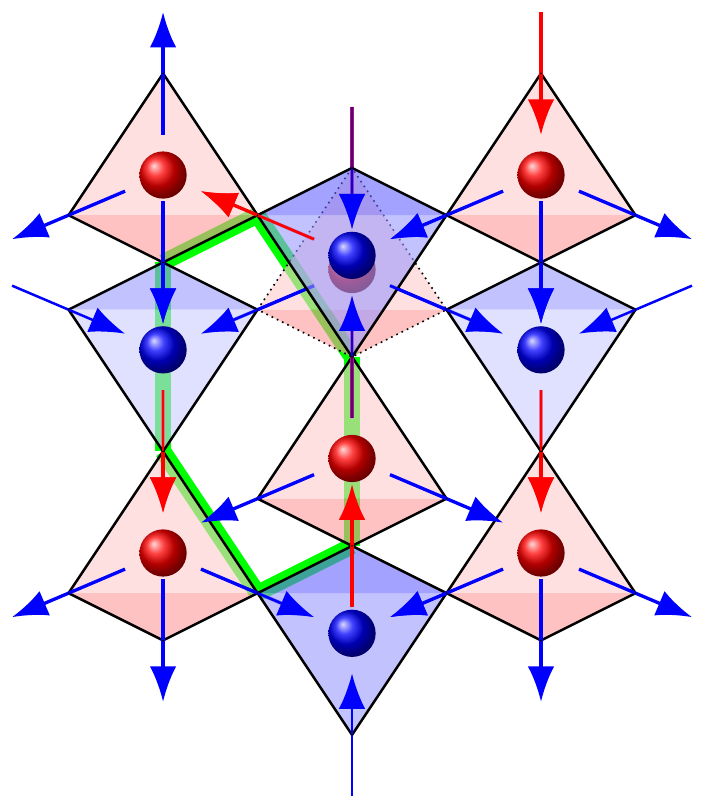}
	\caption{}
\end{subfigure}
\begin{subfigure}[b]{0.22\textwidth}
	\includegraphics[width=\linewidth]{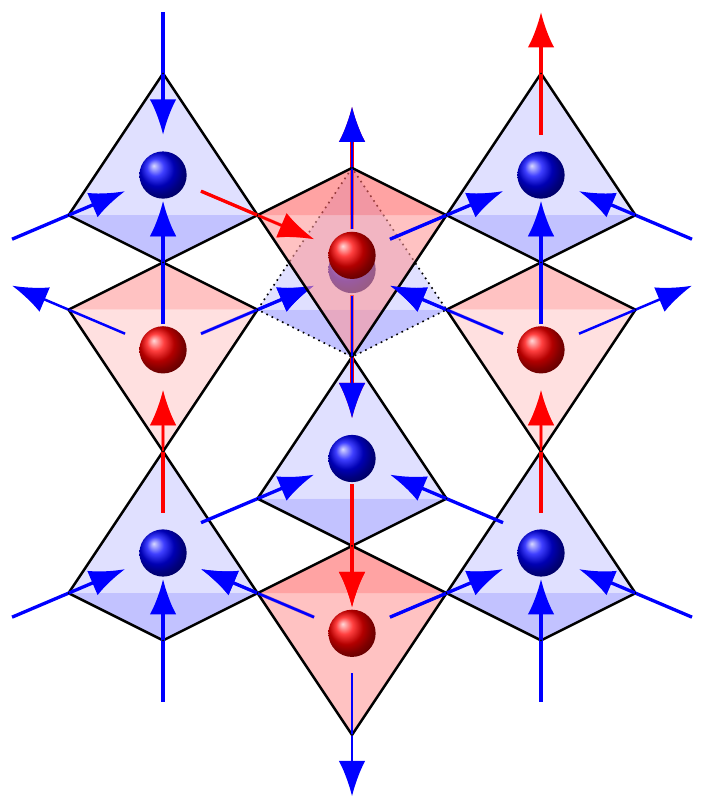}
	\caption{}
\end{subfigure}
\begin{subfigure}[b]{0.45\textwidth}
	\includegraphics[width=\columnwidth]{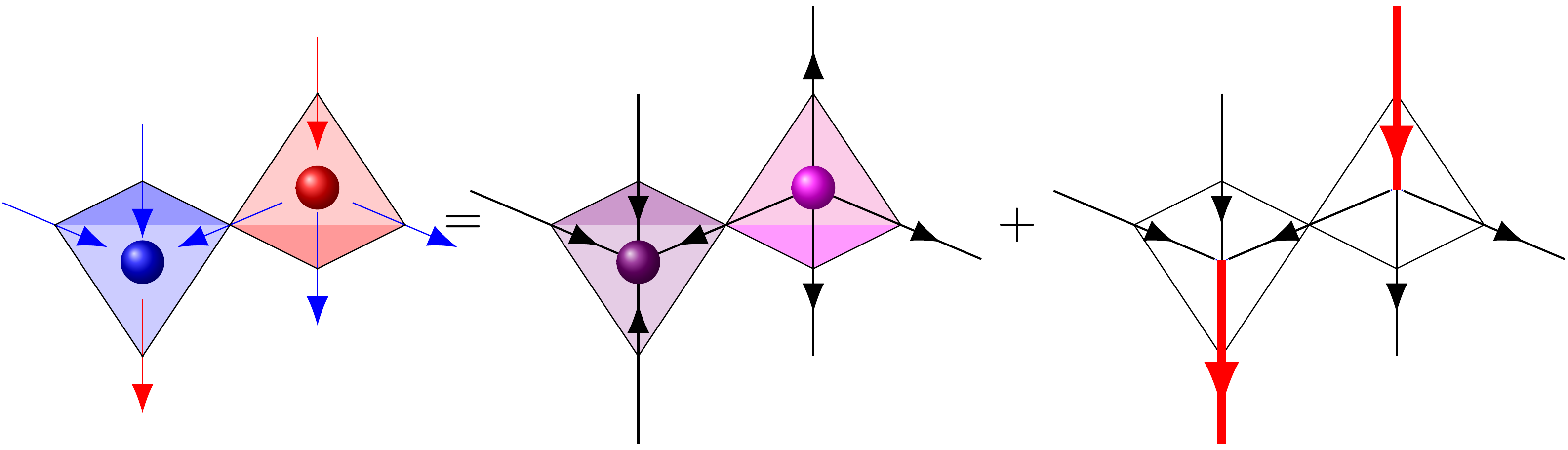}
	\caption{}
	\includegraphics[width=0.3\columnwidth,angle=90]{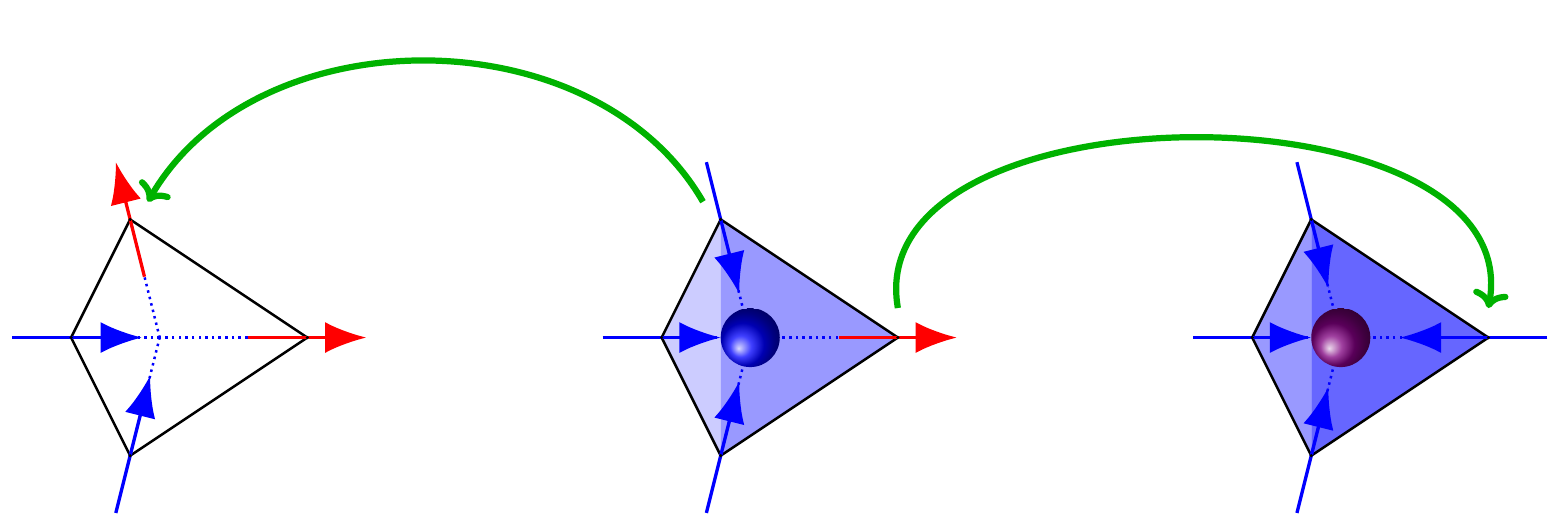}
	\caption{}
\end{subfigure}
\caption{
\textbf{Fragmentation in spin ice and domain walls:} 
The pyrochlore lattice is made of corner-sharing tetrahedra with four spin sublattices, one at each tetrahedron corner. The centres of all tetrahedra form a bipartite diamond lattice.
(a) In a fragmented Coulomb phase, all tetrahedra are for example in a 3in-1out state on up tetrahedra (blue) and a 3out-1in state on down tetrahedra (red), defined as domain $\mathcal{D}$. Flipping all spins around a closed loop of in-out-in-out... spins preserves the domain $\mathcal{D}$ (see green hexagon)
(b) Flipping all spins in the system corresponds to domain $\mathcal{\bar D}$.
(c) The magnetic degrees of freedom of a 3in-1out state can be decomposed into two parts~\cite{brooks14a}: (i) a 4in, divergence-full, part and (ii) a zero-divergence part whose minority spin (red) is three times bigger that the other ones. The minority spin is shown in red and is always shared between two adjacent tetrahedra. Once applied to the entire lattice, the former part (i) is responsible for a partial all-in/all-out (AIAO) antiferromagnetic order, while the latter part (ii) maps exactly onto a hard-core dimer model on the diamond lattice with extensive degeneracy~\cite{nagle66c}.
(d) Starting from a ground state with 3in-1out (centre), flipping one spin creates an excitation. If the flipped spin is a minority spin, then it creates a pair of 4in and 4out defects (right). If the flipped spin is a majority spin, then it creates a pair of 2in-2out defects (left). In both cases, excitations out of the fragmented phase are topological charges, created in pairs (one positive and one negative). They interact via an entropic Coulomb potential, in analogy with traditional spin-ice monopoles~\cite{jaubert15c}.
}
\label{fig1}
\end{figure*}

Besides these practical concerns, fragmentation offers a framework for fundamental questions in emergent phenomena. 
We know that spin ice supports an emergent Coulomb gauge field and magnetic monopoles~\cite{castelnovo08a,Henley10a}. 
From this point of view, spin ice can be seen as an artificial ``world'', whose low-temperature properties are governed by a simplified version of Maxwell's electromagnetism on a lattice. 
Fragmentation couples this emergent electromagnetism with an underlying broken symmetry, namely time-reversal symmetry due to the presence of magnetic order. 
A domain wall in fragmented spin ice thus separates one ``world'' from its time-reversal symmetry related partner. Quasi-particles can a priori move from one side to the other, raising the natural question of what happens at the crossing.\\

In this paper, we first introduce the notion of fragmentation. 
Then we present how, for certain geometries, domain walls can filter positive from negative monopoles. 
This mechanism is confirmed by simulations, making domain walls the magnetic analogues of a diode for monopole current. 
Keeping in mind our motivation to design devices for magnetricity, we show that successive domain walls enable the accumulation of charges of opposite signs on opposite sides of the system, realising a capacitor for magnetic carriers. 
Then we discuss how domain wall fluctuations are able to effectively change the sign of monopoles. 
Finally, we discuss directions and challenges to realise our theory in experiments.

\section{Fragmented spin ice}
\label{secFSI}

In spin-ice materials, rare-earth magnetic ions, Dy$^{3+}$ or Ho$^{3+}$, form a pyrochlore lattice~\cite{Harris97a}. Their magnetic moments have a strong local easy-axis anisotropy, behaving like Ising spins pointing either towards, or away from, the centre of each tetrahedron. Upon cooling, spin-ice compounds famously remain magnetically disordered~\cite{spinicebook}. Their ground state is a vacuum of monopoles called a Coulomb phase~\cite{Henley10a}, where each tetrahedron has two spins pointing in, and two spins pointing out (2in-2out). It implies that all the magnetisation flux entering a tetrahedron exits the tetrahedron as well, and the Coulomb phase state is characterised by a conserved magnetisation flux with zero divergence, a lattice analogue of Maxwellian magnetostatics. Monopoles are point-like excitations with a topological charge that breaks the zero divergence. Hence, they correspond to sources or sinks of this magnetisation flux. It is this correspondence that confers an effective magnetic charge to monopoles, in addition to their topological charge.

\begin{figure*}[t]
\centering
\begin{subfigure}[b]{0.9\columnwidth}
	\includegraphics[width=\columnwidth]{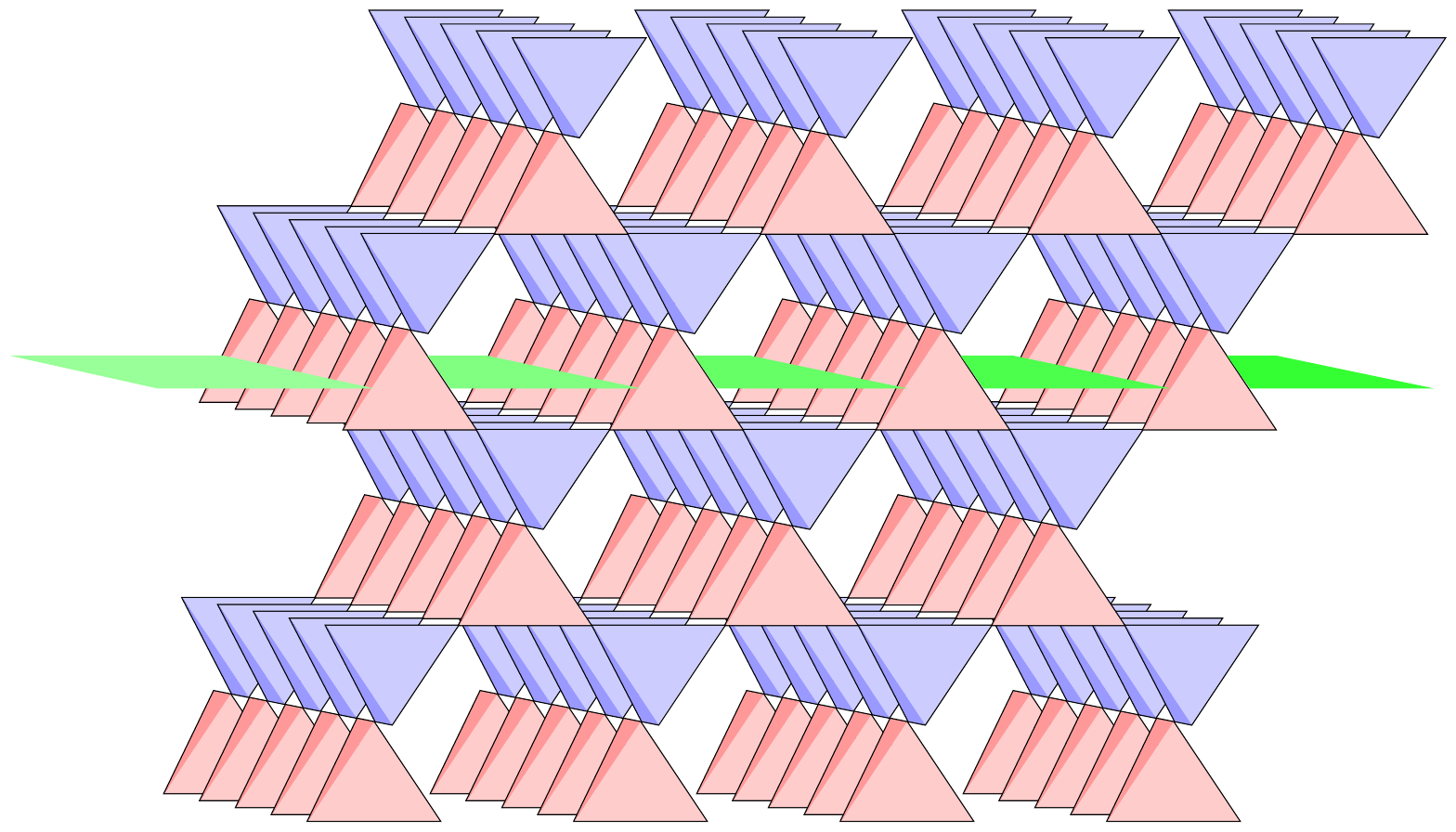}\qquad
	\caption{}
\end{subfigure}
\begin{subfigure}[b]{0.9\columnwidth}
	\includegraphics[width=\columnwidth]{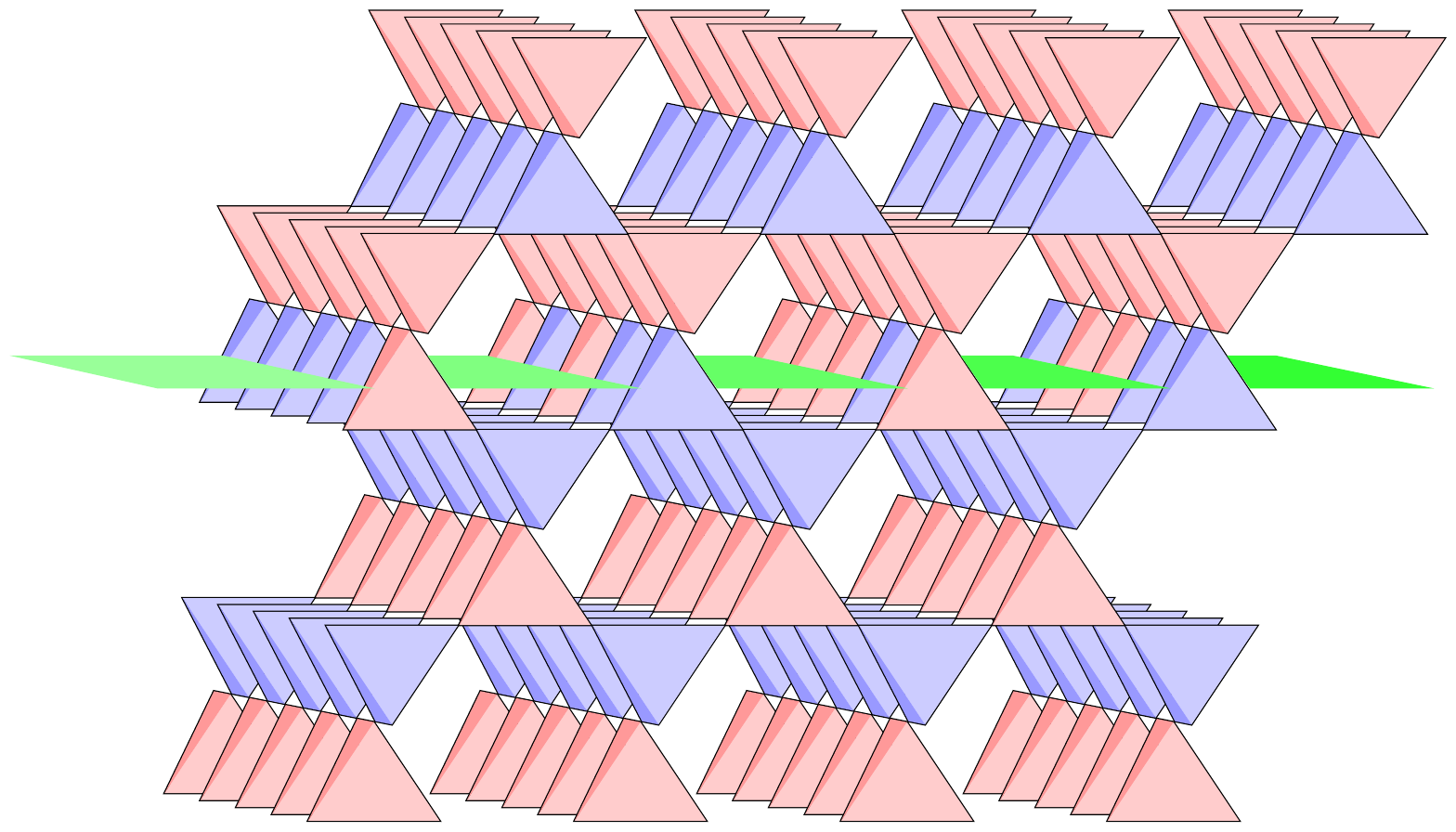}
	\caption{}
\end{subfigure}
\caption{
\textbf{Domain walls in fragmented spin ice:}
(a) Starting from a fragmented spin configuration in domain $\mathcal{D}$, we define a (green) plane at layer $z=n_{dw}$, orthogonal to the [001] direction.
(b) Flipping all spins above this plane transforms the upper part of the system from domain $\mathcal{D}$ to $\mathcal{\bar D}$. The domain wall is one-tetrahedron layer thick, where 3out-1in and 3in-1out states co-exist; the former (red) belong to the lower domain $\mathcal{D}$, while the latter (blue) belong to upper domain $\mathcal{\bar D}$. 
}
\label{fig2}
\end{figure*}

There are, however, many forms of Coulomb phases, some of them hidden behind a type of Helmholtz decomposition known as fragmentation~\cite{brooks14a}. Fragmentation essentially consists in dividing the magnetisation flux into its zero-divergence and divergence-full parts [Figure \ref{fig1}(c)]. The latter component crystallises into long-range antiferromagnetic order (known as all-in/all-out (AIAO), see Figure \ref{fig1}(c)), while the remaining magnetic degrees of freedom remain fluctuating with zero-divergence flux and form an extensively degenerate Coulomb phase. In magnetic fragmentation, the Coulomb spin liquid and AIAO order are thus coupled together in an integral way. Fragmented phases have been confirmed theoretically in a variety of microscopic models~\cite{moller09a,chern11a,borzi13a,brooks14a,jaubert15c,lefrancois17a,Raban19a,slobinsky21a,Palle21a,Szabo22a,museur23a,saccone23a,Vijayvargia23a,Yogendra24a,Rougemaille25a,lhotel20a} and observed experimentally in Iridium- and Ruthenium-based pyrochlore oxides~\cite{lefrancois17a,Cathelin20a,vlaskova21a,Pearce22a,museur26a}, in various setups of artificial spin ice~\cite{Canals16a,Sendetskyi16a,Yue22a,saccone23a,rougemaille19a,Parakkat19a}, in a series of 2D and 3D materials~\cite{Petit16a,benton16b,Xu20a,paddison16a,dun20a,lhotel20a} as well as proposed in spin-crossover materials~\cite{Cruddas21a}.

\begin{table}[h!]
\centering
\begin{tabular}{|c|c|c|}
\hline
Ground state & Positive charge $\oplus$ & Negative charge $\ominus$  \\
\hline
3in-1out & 4in & 2in-2out \\
3out-1in & 2in-2out & 4out \\
\hline
\end{tabular}
\caption{
\textbf{Charge excitations in fragmented spin ice:} for a given ground state, flipping an inward (resp. outward) spin creates a negative (resp. positive) charge~\cite{castelnovo08a,brooks14a}.
}
\label{tab1}
\end{table}

In a typical fragmented state on pyrochlore, all up tetrahedra are 3in-1out, while all down tetrahedra are 3out-1in (or vice-versa) [Figure \ref{fig1}(a)]. 
Once time-reversal symmetry is broken by the AIAO order, the spin configuration maps exactly onto a hard-core dimer model on the diamond lattice (the Coulomb phase)~\cite{brooks14a}, where the minority spin of each tetrahedron is the dimer. 
From the point of view of traditional spin ice, fragmentation can be seen as a crystal of monopoles in a zinc-blende structure. But this magnetic texture is now the ground state, and excitations take the form of 2in-2out, 4in and 4out states [Figure \ref{fig1}(d)]. 
These point-like excitations are the new topological monopoles of the system, with emergent entropic Coulomb correlations, and where magnetic dipolar interactions are rewritten as an effective Coulomb potential, conferring a magnetic charge to the topological one~\cite{brooks14a,jaubert15c}. 
The concept of magnetricity thus directly applies to fragmentation.
The definition of positive and negative charges is more subtle than in spin ice though, as it can only be defined with respect to the local AIAO order, as summarised in table \ref{tab1}. 
Since an excitation must conserve its topological charge over time, a charge hopping on the lattice thus takes a different form \emph{depending} on whether it is on an up or a down tetrahedron~\cite{jaubert15c,lefrancois17a}: e.g. a positive charge would be successively 4in, then 2in-2out, then 4in, then 2in-2out and so on.

To conclude, we shall emphasise that there are many different models able to stabilise a fragmented phase. 
But in this paper, we wish to develop a general theory of domain walls for magnetic fragmentation, independent of those microscopic details. 
In other words, we will focus on the geometrical constraints imposed by domain walls, and consider a minimal model with random monopole hopping~\cite{castelnovo11a,jaubert11c,jaubert15c}. 
For that reason, we will consider that 4in, 4out and 2in-2out excitations have the same chemical potential, as it can in practice vary one way or the other in materials~\cite{lefrancois17a}. 
We do not expect this choice to affect our results, since the charge always comes back to the same state every two steps. 
Following the discussion on monopole hopping of the previous paragraph, any energy cost due to a different chemical potential at a given step is recovered at the next hopping, and the dynamics thus remains deconfined.

\section{Results}
\subsection{Domain walls}
\label{secDW}

As discussed in the previous sections, a fragmented phase is partially ordered magnetically, with a $\mathbb{Z}_2$ order parameter of the Ising Universality class. 
And it is this broken time-reversal symmetry that offers a new prospect absent from traditional spin ice; the domain walls. 
Locally, the Coulomb phase couples to a given antiferromagnetic domain, but moving across a domain wall means flipping its entire background.
Despite their experimental observations in Dy$_2$Ir$_2$O$_7$~\cite{Cathelin20a}, Ho$_2$Ir$_2$O$_7$~\cite{Pearce22a} and Ho$_2$Ru$_2$O$_7$~\cite{museur26a}, and their potential magneto-electric activity~\cite{Khomskii21a}, little is known about the influence of these domain walls. Besides the fundamental question of what happens at this frontier between time-reversal symmetric Coulomb phases, their potential applications to the control of monopole transport is an open question.

But first, what does a domain wall look like ? Let us define that domains $\mathcal{D}$ and $\mathcal{\bar D}$ of the fragmented phase respectively correspond to all up tetrahedra either in a 3in-1out, or in a 3out-1in state [Figure \ref{fig1}(a,b)]. Within a given domain, any layer of tetrahedra orthogonal to the [001] axis is made of the same state, whose sign alternates between adjacent layers [Figure \ref{fig2}(a)]. Let us assume arbitrarily that our system is initially in domain $\mathcal{D}$, and choose a (001) plane crossing the layer of tetrahedra at height $z= n_{dw}$. By flipping all spins above this plane [Figure \ref{fig2}(b)], layers at $z<n_{dw}$ remains in domain $\mathcal{D}$, while layers at $z>n_{dw}$ turn into domain $\mathcal{\bar D}$. For each tetrahedron at layer $z=n_{dw}$, two spins have been flipped. If these two spins were 1in-1out, then these two spins become 1out-1in after flipping, and the tetrahedron remains in the same state, and thus in domain $\mathcal{D}$. If not, then the two spins were initially either 2in or 2out. After flipping, they have turned into 2out or 2in, respectively; the tetrahedron now belongs to domain $\mathcal{\bar D}$. This specific magnetic texture at the interface between the two domains will play an important role in the following. The domain wall is almost flat, with a roughness of one-layer thickness. Note that there are a priori two kinds of domain walls; those going through a layer that was initially either 3in-1out or 3out-1in. We will soon elaborate on this concept, but at this stage, we shall refer to these domain walls as $\mathcal{W}_-$ and $\mathcal{W}_+$ respectively.

\begin{figure*}
\centering
\centering
\begin{subfigure}[b]{0.455\columnwidth}
	\includegraphics[width=\columnwidth]{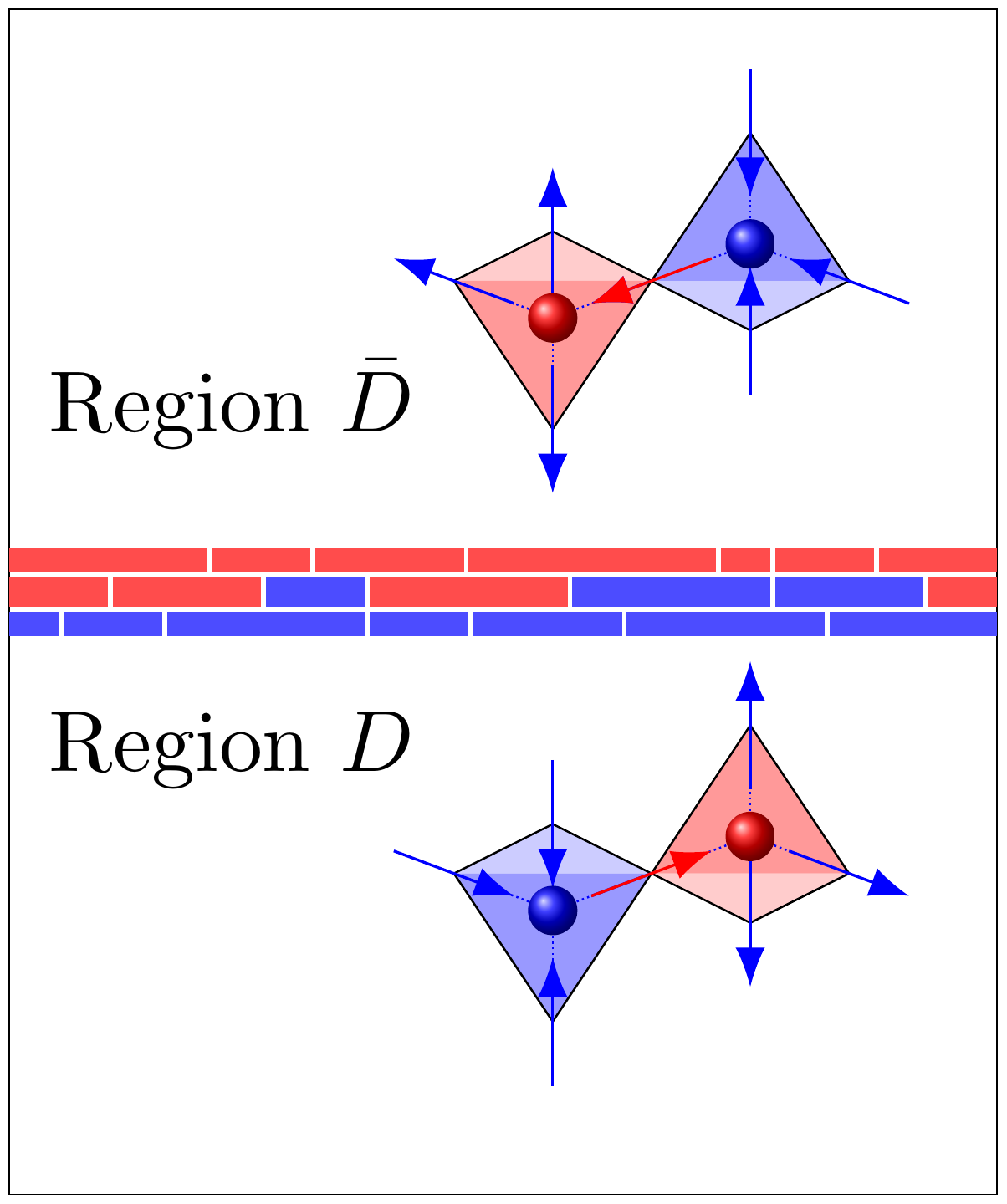}
	\caption{}
\end{subfigure}
\begin{subfigure}[b]{0.72\columnwidth}
	\includegraphics[width=\columnwidth]{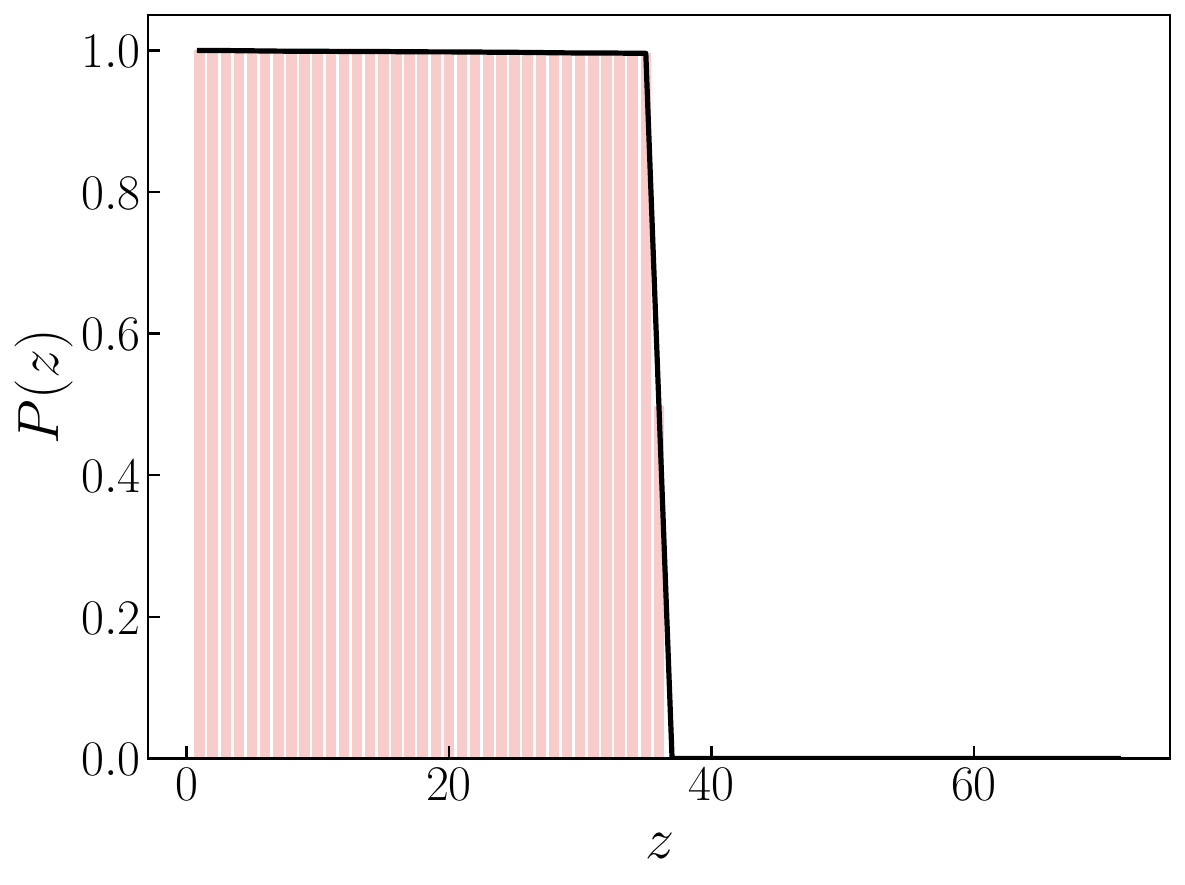}
	\caption{}
\end{subfigure}
\begin{subfigure}[b]{0.72\columnwidth}
	\includegraphics[width=\columnwidth]{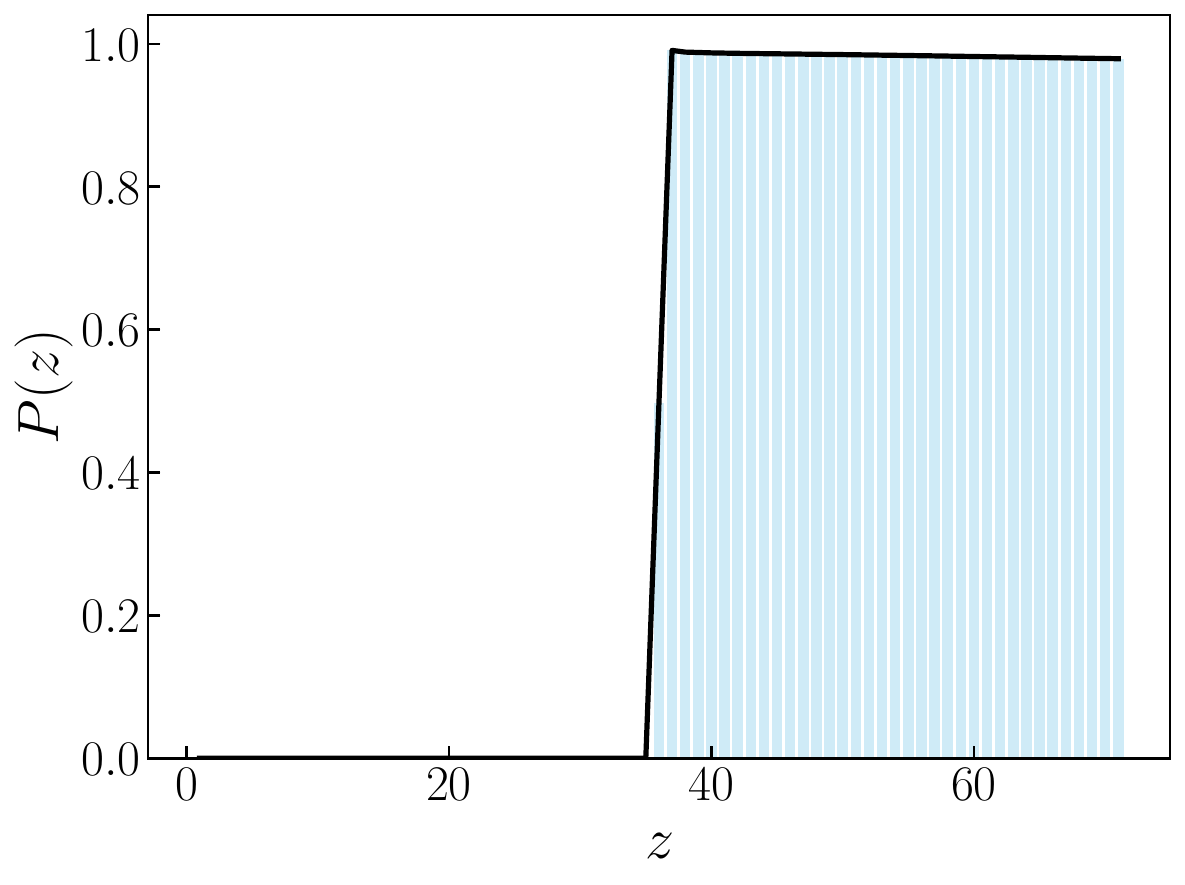}\\
	\caption{}
\end{subfigure}
\begin{subfigure}[b]{0.4\linewidth}
	\includegraphics[width=\linewidth]{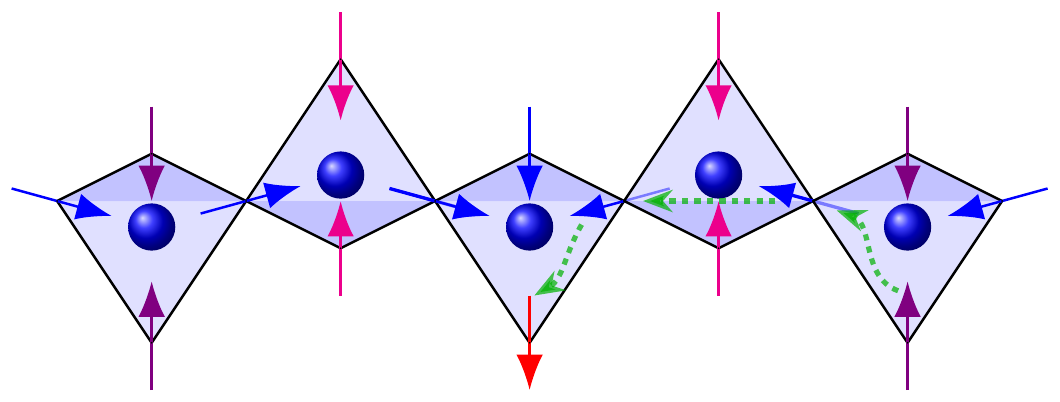}
	\caption{}
\end{subfigure}\qquad\qquad\qquad
\begin{subfigure}[b]{0.4\linewidth}
	\includegraphics[width=\linewidth]{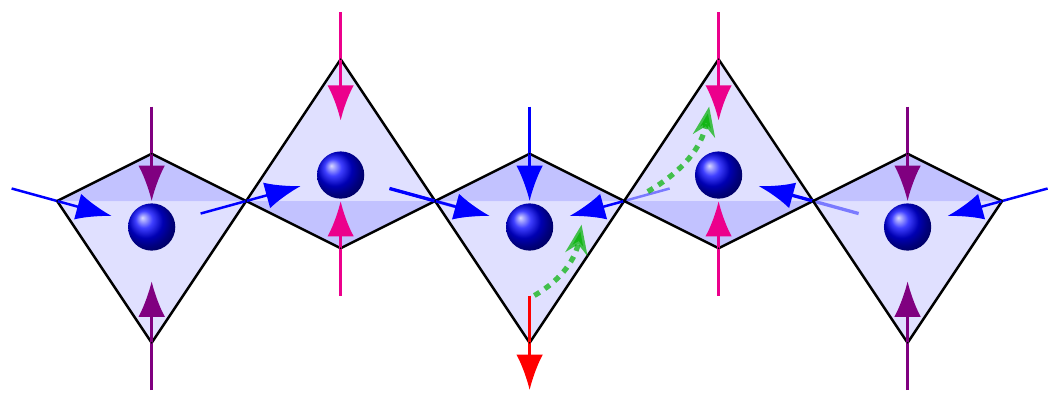}
	\caption{}
\end{subfigure}
\caption{
\textbf{Diode effect:}
(a) Schematic representation of the system divided into domains $\mathcal{D}$ (bottom) and $\mathcal{\bar D}$ (top) via a domain wall at $z=n_{dw}=L/2$.
(b,c) Probability distribution function of the position of monopole carriers, as a function of $z$, during their diffusion in presence of a domain wall, for negative (b) and positive (c) charges. 
Simulation details are given in appendix \ref{appMC}.
(d,e) Microscopic mechanism of the diode effect. Starting from a given initial state at the domain wall, the green paths illustrates how (d) a negative charge is reflected, while (e) a positive charge can be transmitted. Coming from below, this interface is a $\mathcal{W}_+$ domain wall.
}
\label{fig3}
\end{figure*}

\subsection{Diode for magnetic monopoles}

It is the domain wall that will serve as a diode for monopole current. At this stage, the analogy remains somewhat obscure, and the rest of this section will be devoted to clarify it. However, the underlying reasoning is already comprehensible. Fragmentation breaks time-reversal symmetry via its $\mathbb{Z}_2$ antiferromagnetic order, introducing an ``arrow of time''. By breaking a spatial mirror symmetry, the domain wall couples this arrow of time with a spatial direction (here the [001] axis). This coupling makes, at least conceptually, possible to favour opposite directions for the transport of magnetic carriers of opposite charges. It also means that, once the domain wall is established, this effect does not require any external magnetic field since time-reversal symmetry is already inherently broken. This is an important property for a potential diode, in the same sense that the depletion region of a $p$-$n$ junction is present in absence of any electric current or external potential.

Our goal now is to show that the domain wall plays the role of an asymmetric filter, allowing monopoles to cross in one direction and to be reflected in the opposite direction, depending on their charge.

\subsubsection{Simulations}

In order to build our analysis on an unbiased method, we used the numerical protocol that has been validated in Refs.~\cite{castelnovo11a,jaubert11c,jaubert15c} when confirming the Debye-H\"uckel theory and Coulomb potential between magnetic monopoles in spin ice. Here we adapted this protocol to the presence of domain walls in the fragmented phase.

To initiate the simulation, we shall create a pair of monopoles of opposite charge in one domain, let one of them hop around randomly, until it hits its opposite partner, and observe the effect of the domain wall depending on the sign of the hopping charge [Figure \ref{fig3}(a)]. 
For the time being, we only consider \textit{one} pair of monopoles. A higher density of carriers will be explored in following sections. 
The present simulations allow for low-energy dynamics, as if the hopping monopole was essentially alone. 
Monopole creation is forbidden, while monopole annihilation ends the simulation. 
Also, we consider an \textit{immobile} domain wall; the effects of fluctuations will be discussed in Sec.~\ref{sec:disc}. 
Monopoles can go through the domain wall, but they cannot move it. 
It means that any given tetrahedron always remains part of the same domain, either $\mathcal{D}$ or $\mathcal{\bar D}$, throughout the monopole dynamics. 
Spins are allowed to flip at the domain wall, as long as the interface remains fixed. 
The possibility of spin fluctuations while preserving the immobility of the interface is a direct consequence of the extensive degeneracy of fragmented spin ice. 
For efficiency purposes, we prevent backtracking. 
The monopole hops randomly on the lattice, but it cannot immediately hop back onto the previous tetrahedron. 
The main advantage is to hinder annihilation at very short time scales, since neighbouring monopoles cannot immediately annihilate each other without backtracking. 
Instead, the shortest path for annihilation is a loop of six spins. 
This is a pragmatic choice to effectively induces a short, but finite, distance between monopoles at the beginning of simulations. 
Data is obtained by averaging over many independent simulation runs. All details of the simulation are given in appendix \ref{appMC}.

Figure \ref{fig3}(b,c) presents the probability for a monopole of a given charge, either (b) negative or (c) positive, to be at layer $z$. 
The distributions are essentially flat in one of the two domains, and abruptly vanish in the other one. 
Negative charges strictly stay in domain $\mathcal{D}$ ($z<n_{dw}$), while positive ones can cross the domain wall $\mathcal{W}_+$ into domain $\mathcal{\bar D}$ ($z>n_{dw}$) where they are confined. 
The mechanism behind this will be explained soon in the following section.
Domain walls are thus hard interfaces, imposing either reflection or transmission depending on the charge \emph{and} on the domain, but have negligible effect once the charge has stepped away from them since the distributions are essentially flat. 
This is precisely the type of asymmetric filtering we are looking for in a diode.

\subsubsection{Mechanism}
\label{sec:mech}

Before creating the domain wall, when the entire system is in domain $\mathcal{D}$, let us arbitrarily choose a layer where all tetrahedra are in a 3out-1in state. This will be the layer of our domain wall $z=n_{dw}$, and the two layers at $z=n_{dw}\pm 1$ are thus initially 3in-1out [Figure \ref{fig2}(a)]. Once the $\mathcal{W}_+$ domain wall is formed at layer $z=n_{dw}$, tetrahedra at $z=n_{dw}-1$ remain 3in-1out (part of domain $\mathcal{D}$), but the ones at $z=n_{dw}+1$ have flipped and are now 3out-1in (part of domain $\mathcal{\bar D}$) [Figure \ref{fig2}(b)]. As for tetrahedra at $z=n_{dw}$, the situation is more nuanced as discussed in section \ref{secDW}. They remain in domain $\mathcal{D}$ if their configuration has been preserved, i.e. if the two top spins that have been flipped were 1in-1out. On the opposite hand, if the two top spins were 2out, they have now been flipped into 2in; the tetrahedron is now 3in-1out and belongs to domain $\mathcal{\bar D}$. 

This subtle difference at layer $z=n_{dw}$ is the key for the asymmetric filtering of the domain wall. During its hopping on the lattice, we know that a negative charge will alternatively take the form: 4out $\rightarrow$ 2in-2out $\rightarrow$ 4out $\rightarrow$ 2in-2out ... (see section \ref{secFSI}). A negative charge always enters (resp. exits) a tetrahedron through an inward (resp. outward) spin of that given tetrahedron. It remains true at the domain wall as long as the domain wall is immobile. In order to cross the $\mathcal{W}_+$ interface from below (i.e. from domain $\mathcal{D}$), a negative charge thus needs to access layer $n_{dw}$ via an inward spin. But since we know that the 3out-1in tetrahedra of layer $n_{dw}$ necessarily have their inward spin at the top (see above discussion and Figure \ref{fig3}(d)), this entry point for a negative charge is not accessible from below. A negative charge can thus only enter layer $n_{dw}$ through a 3in-1out tetrahedron, which provides only one outward spin for exit. In order to cross the $\mathcal{W}_+$ domain wall (and move from layer $n_{dw}$ to $n_{dw}+1$), this exit point needs to be at the top of the tetrahedron. However, we know this is impossible since the two top spins are 2in (see above discussion). It means the charge necessarily has to come back to layer $n_{dw}-1$. From there, it can either move deeper into domain $\mathcal{D}$ ($z<n_{dw}-1$), or move back to the interface at $n_{dw}$. But the same argument will then apply; a negative charge can only enter layer $n_{dw}$ through tetrahedra which forbid access to higher layers. The negative carrier will thus eventually be reflected by the $\mathcal{W}_+$ domain wall and diffuse back into domain $\mathcal{D}$.

As illustrated in Figure \ref{fig3}(e), the same line of reasoning can be applied to a positive charge, but with a different outcome. A positive charge always enters (resp. exits) a tetrahedron through an outward (resp. inward) spin of that given tetrahedron. Hence, tetrahedra of layer $n_{dw}$ that have stayed in a 3out-1in state can always be entered by a positive charge through an outward spin from below, and can only exit by an inward spin situated at the top of the tetrahedron into layer $n_{dw}+1$. As for tetrahedra of layer $n_{dw}$ that have flipped into 3in-1out, both transmission and reflection are possible. By time-reversal symmetry, the exact opposite applies to a $\mathcal{W}_-$ domain wall, i.e. in a layer that was initially 3in-1out, as summarised in Table \ref{tab2}.

\begin{table}[h!]
\centering
\begin{tabular}{|c|c|c|}
\hline
Charge carrier & Domain-wall facet & Diode effect \\
\hline
$\ominus$ & $\mathcal{W}_+$ & Reflection \\
$\oplus$    & $\mathcal{W}_+$ & Transmission \\
\hline
$\ominus$ & $\mathcal{W}_-$ & Transmission \\
$\oplus$    & $\mathcal{W}_-$ & Reflection \\
\hline
\end{tabular}
\caption{
\textbf{Diode effect} for positive ($\oplus$) and negative ($\ominus$) charge carriers depending on the facet of the domain wall. While reflection is strictly imposed, transmission is not compulsory; it is the most probable process, but reflection is also allowed.
}
\label{tab2}
\end{table}

But then what happens once the monopole has crossed ? Let us come back to our positive charge in domain $\mathcal{D}$ that crosses a $\mathcal{W}_+$ domain wall from below before entering domain $\mathcal{\bar D}$. When this charge hits again the same interface from the top, i.e. from domain $\mathcal{\bar D}$, it will now see a $\mathcal{W}_-$ domain wall, and thus be reflected. This is because domain $\mathcal{\bar D}$ is the mirror image of $\mathcal{D}$. At layer $n_{dw}$, tetrahedra of domain $\mathcal{D}$ are 3out-1in, but tetrahedra from domain $\mathcal{\bar D}$ are 3in-1out. \textit{A domain wall always presents two facets}; if one side is $\mathcal{W}_+$, then the other side is $\mathcal{W}_-$, and vice-versa. If a carrier crosses an interface, it won't be able to cross back from the other side. This is how we obtain charge separation in Figure \ref{fig3}(b,c).

There is only one exception; the original path used by the charge through the wall. Through this path (i.e. through this spin), the positive charge is not reflected and is allowed to cross back into domain $\mathcal{D}$. In our simulations, it is the path that enables the two topological charges to meet again eventually and annihilate. However, this path only includes one spin over a cross section that scales like $L^2$. As confirmed by simulations of Figure \ref{fig3}(b,c), this occurence is sufficiently rare to be negligible in the probability distribution function.\\

Now, we can push the analogy further. Let us assume a current of those magnetic charges, with positive and negative carriers going in opposite directions, as expected in an external magnetic field for example. Depending on its orientation, a single domain wall will either block all carriers, or let them all cross. This is the same effect that an electric diode has on electrons and electron holes. In a setup where the domain wall has not been previously crossed by any carriers, the blockade will be total. On the other hand, if the domain wall has been crossed by some carriers coming from the other side, there will be a memory effect and each crossing carrier will have opened a fixed path through the wall that can be used by one and only one carrier to cross back. In an interesting analogy to electronics, it would correspond to a leakage current, that can be minimised by working at low density of carriers compared to the cross section of the diode.

\begin{figure*}[htbp]
\centering
\begin{subfigure}[c]{0.16\linewidth}
    \centering
    \makebox[\linewidth][c]{%
        \includegraphics[width=1.5\linewidth]{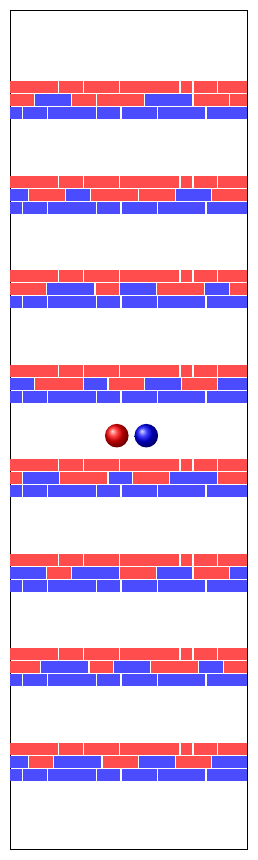}%
    }
    \caption{}
\end{subfigure}
\begin{subfigure}[c]{0.83\linewidth}
	\includegraphics[width=0.6\linewidth]{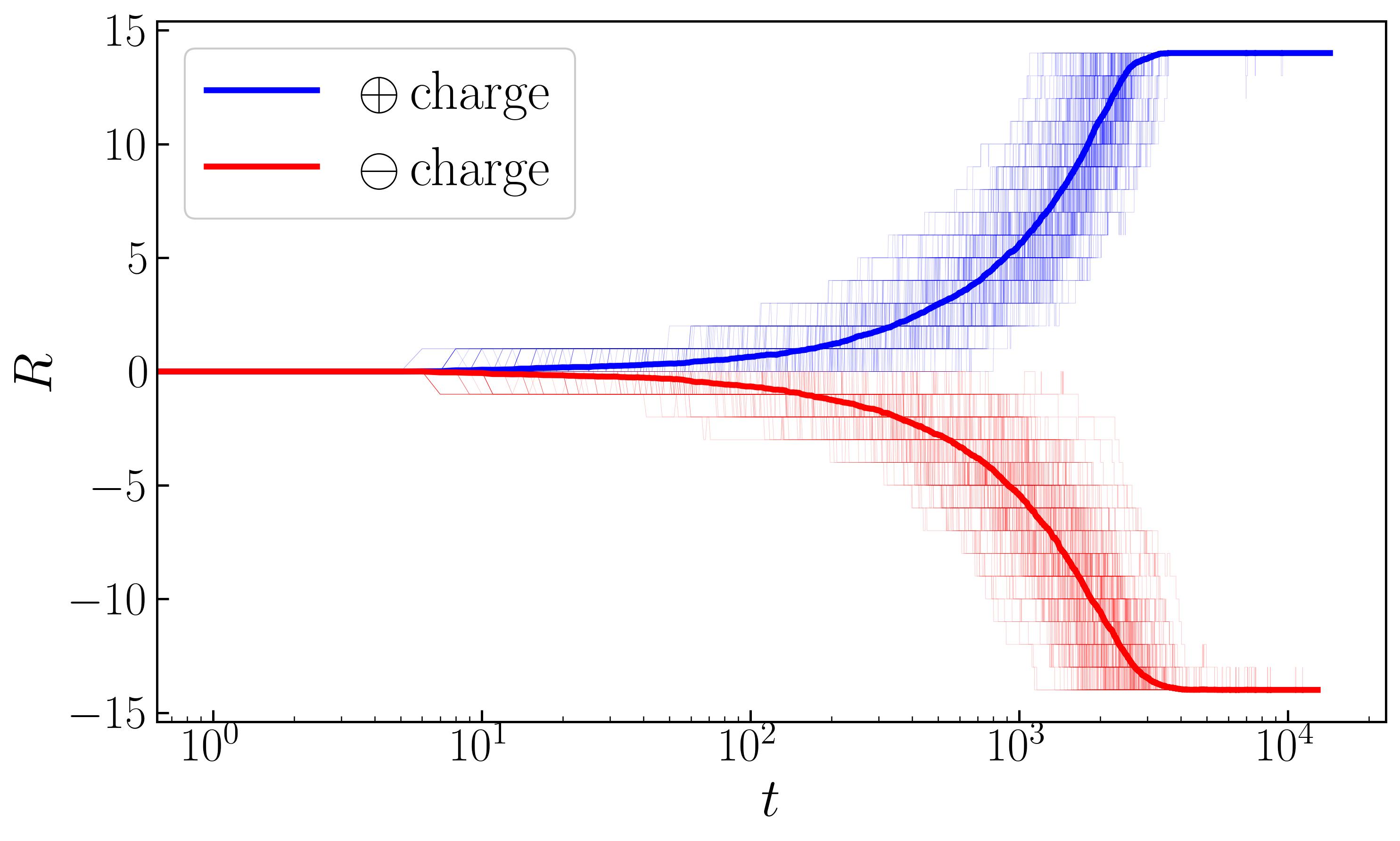}
    \caption{}
	\includegraphics[width=0.6\linewidth]{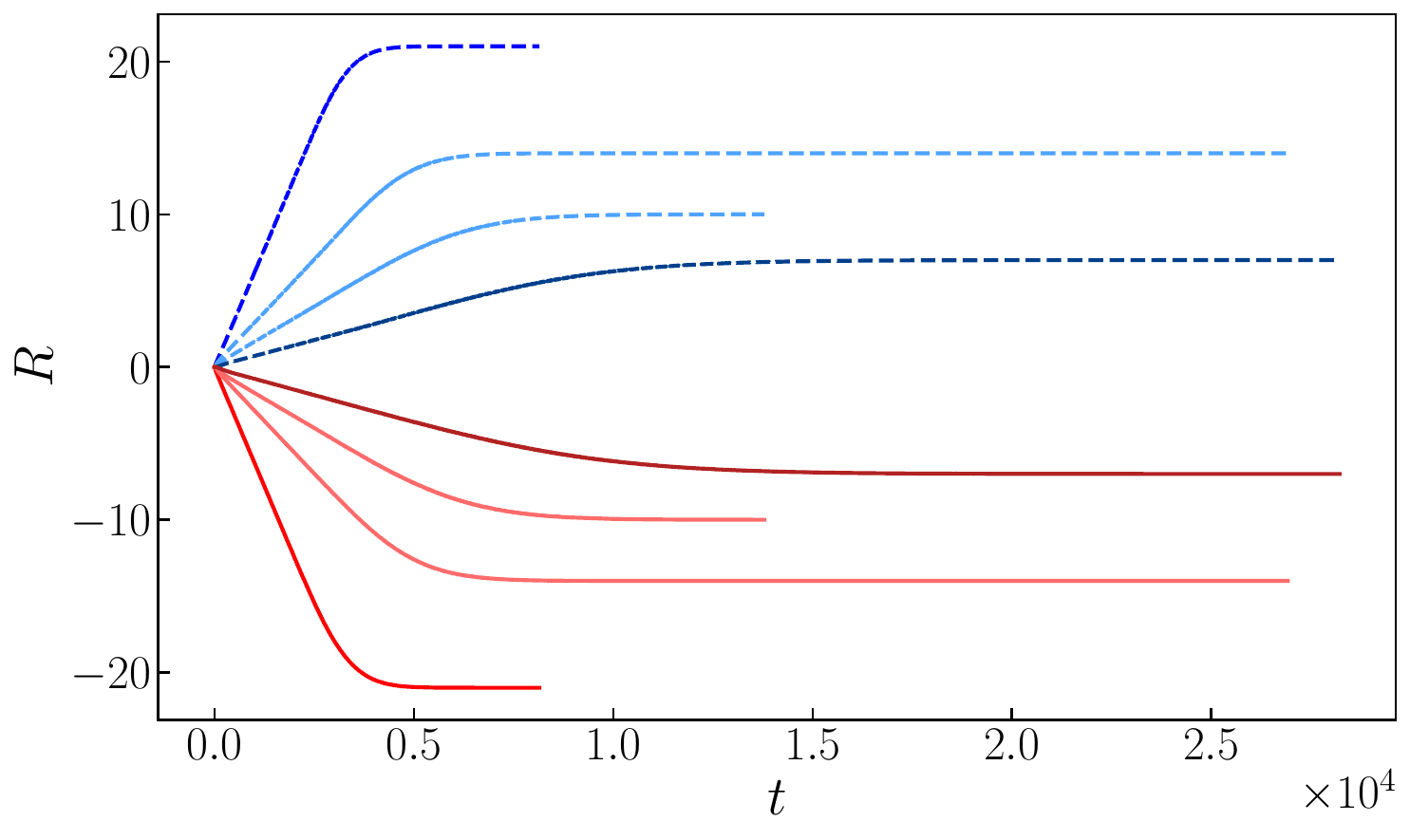}
    \caption{}
	\includegraphics[width=0.6\linewidth]{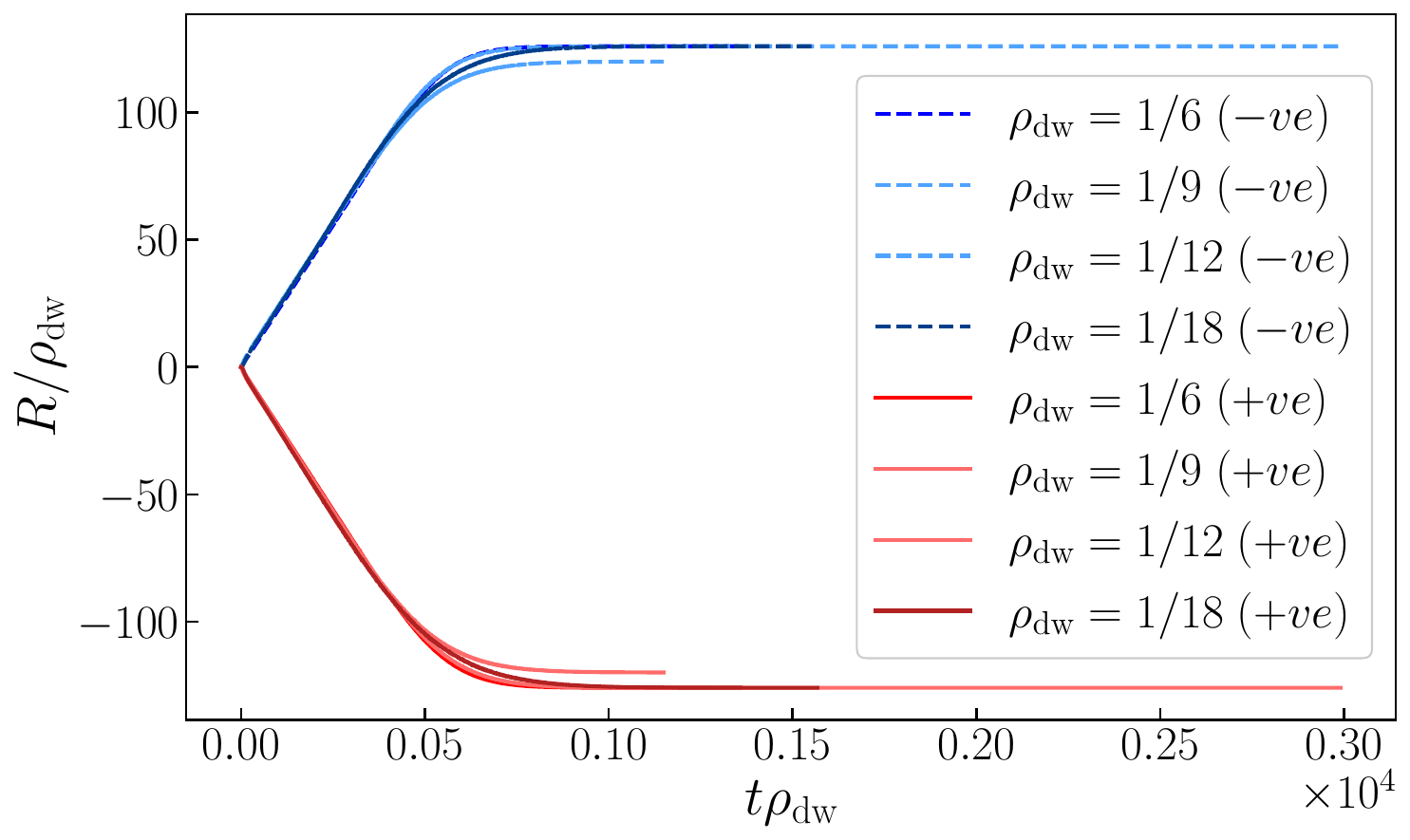}
    \caption{}
\end{subfigure}
\caption{
\textbf{Capacitor:}
(a) Schematic representation of the system with $p$ domain walls, ensuring that positive charges can only go up while negative ones can only go down. A single pair of charges is created in the central region, and its evolution is tracked under Monte Carlo dynamics. 
(b) 
Distribution of the position of positive (blue) and negative (red) charges along the $z-$axis, as a function of simulation time for $p=28$ domain walls and $L=256$. Solid lines represent the average over all independent runs, while thin light-coloured ones  correspond to individual runs. The vertical axis denotes the domain index $i=\{-14, ... , 0 , ... , +14\}$. The horizontal axis is shown on a logarithmic scale. At long times, the charges reach opposite extremities of the system. The leakage current produces rare-event fluctuations into adjacent domains, leading to occasional oscillations into domains $\pm 13$ at late times. 
(c) Same figure for different densities of domain walls $\rho_{dw} = \frac{p}{L_z} = \frac{1}{6}, \frac{1}{9}, \frac{1}{12}, \frac{1}{18}$ and $L=L_z=256$. The horizontal axis is on a linear scale. The initial linear behaviour confirms that carriers possess an effective velocity in this capacitor, despite their random dynamics. As $\rho_{dw}$ increases, so does the velocity; charges are faster to reach the system boundaries. 
The data collapse of (d) confirms the scaling behaviour of Eq.~(\ref{eq:v}). Velocity can be controlled by the number of domain walls.
Simulation details are given in Appendix \ref{appMC}.
 }
\label{fig4}
\end{figure*}



\subsection{Capacitor for magnetic monopoles}
\label{sec:capac}

\subsubsection{Multiple domain walls and simulations}

The next step is to use this asymmetric filter made possible because of fragmentation physics to design more complex devices.
In this section, we will now build on this to design a capacitor for magnetic carriers. To do so, let us consider not one, but $p$ successive domain walls, separating $p+1$ domains labeled by an index $i\in\{-p/2, ... , -1, 0, 1, ... , p/2\}$ [Figure \ref{fig4}(a)]. $L_z$ is the system size along the $z$-direction. Each domain wall is positioned with its $\mathcal{W}_+$ facet facing down and its $\mathcal{W}_-$ facet facing up. Within each domain, a negative charge will be reflected by the upper interface while it can cross the lower one, and vice-versa for positive charges. Hence, negative and positive carriers can only go downwards and upwards respectively.

\begin{figure}
\centering
\begin{subfigure}[b]{\columnwidth}
	\includegraphics[width=\columnwidth]{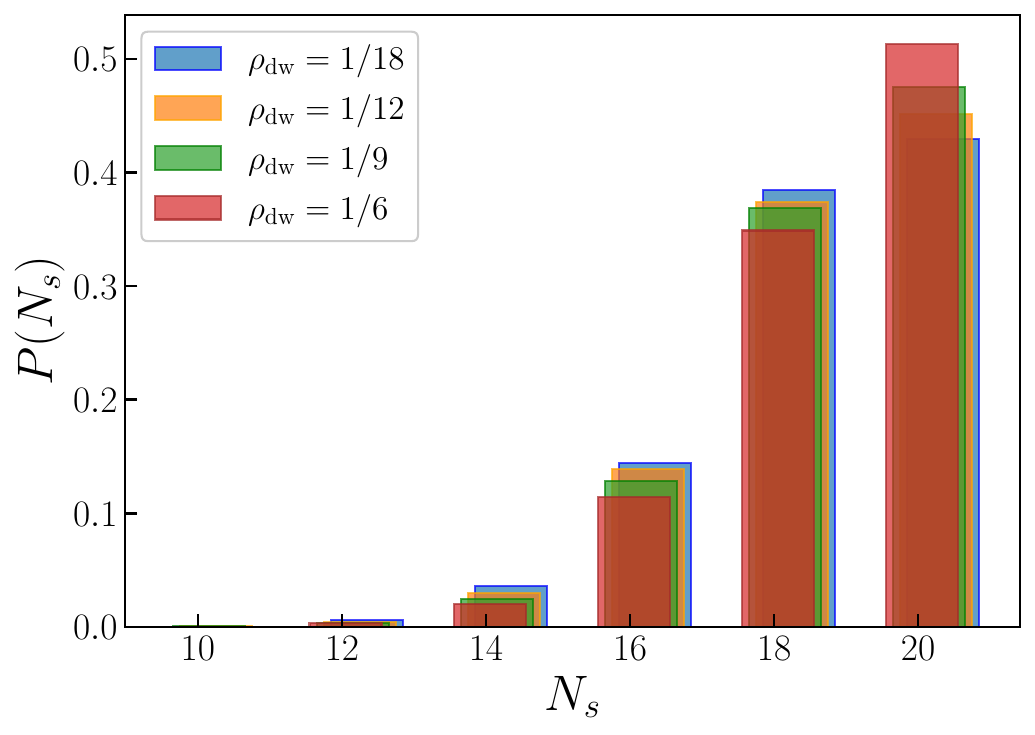}
	\caption{}
	\includegraphics[width=\columnwidth]{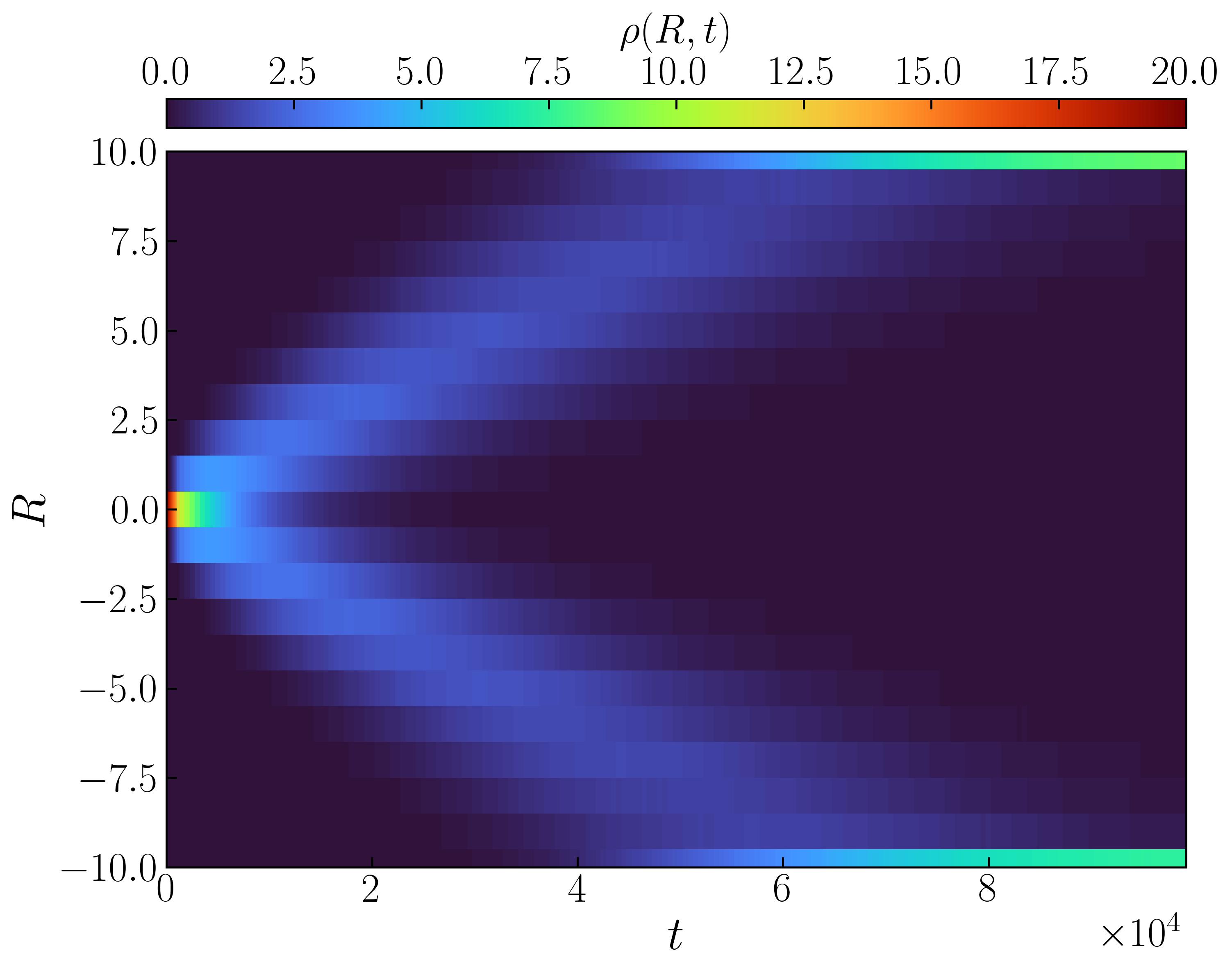}
	\caption{}
\end{subfigure}
\caption{
\textbf{Capacitor with multiple charges:}
Initially, 10 pairs of charges (20 charges) are created in the central (zeroth) region. Annihilation between opposite charges is permitted but domain walls remain immobile. 
(a) Histogram of the number of surviving charges ($N_s$) at the end of simulations, for different densities of domain walls $\rho_{dw} = \frac{1}{6}, \frac{1}{9}, \frac{1}{12}, \frac{1}{18}$ and $L = 256$. 
(b) Space and time evolution of those 20 charges when $p=21$ domain walls. The colour scale represents the average number of surviving charges in a given layer. Upon increasing time, the charges move outward and eventually accumulate near the edges. At the edges, the average number of surviving charges is on average a bit less than 10 (see panel (a)).
Simulation details are given in appendix \ref{appMC}.
}
\label{fig5}
\end{figure}

We test this idea in simulations, creating a pair of charges in the middle of the system, and letting both charges move with the same dynamics as before (see appendix \ref{appMC}). Figure \ref{fig4}(b) shows how the positions of the two charges evolve as a function of time. For a given simulation, the curve necessarily evolves in steps (light-coloured lines) as it crosses from one domain to the next. But once averaged over several independent runs (solid lines), one obtains a smooth curve, symmetric with respect to the sign of the charge. According to Figure \ref{fig4}(c), the propagation of the charge is linear with time, away from the extremities of the system. Despite the random nature of the monopole dynamics, we thus have a quasi-ballistic transport of the carriers with a well defined velocity $v$, similar to the ratchet effect in active matter~\cite{reichhardt17a}. Since all domains approximately have the same thickness $L_z/p$ in our setup, the time necessary to cross each domain is, on average, the same. Assuming a Brownian motion within each domain~\cite{jaubert11a}, this crossing time should scale as $(L_z/p)^2$. Since the carrier needs to cross $p/2$ domains over a distance of $L_z/2$ (half of the system size), its velocity is
\begin{equation}
v\propto\frac{L_z/2}{p/2}\left(\frac{p}{L_z}\right)^2=\frac{p}{L_z}=\rho_{dw},
\label{eq:v}
\end{equation}
proportional to the linear density of domain walls $\rho_{dw}$, as confirmed by the scaling behaviour of Figure \ref{fig4}(d).

\subsubsection{Protocol to charge the capacitor}
\label{sec:protoc}

Let us consider the previous design of Figure \ref{fig4}(a) and assume a very low base temperature with a vanishing density of monopoles. 
If we heat up the central region, thermal fluctuations will induce a local monopole bath. 
In absence of domain walls, the physics is trivial. 
Once the heating stops, the system cools down to the base temperature and all monopoles will eventually annihilate in pairs. 
But in the above setup of Figure \ref{fig4}(a), while the heating induces a sudden bath of monopoles in the central region at $i=0$, some of the thermal monopoles will then cross the domain walls, without possibility to come back. 
With time, a growing number of negative (resp. positive) charges will populate the lower (resp. higher) regions. 
Hence, even if the heating of the central region is stopped, the monopoles which have crossed at least one domain wall will have to continue to the extremities as in Figure \ref{fig4}(b). 
Since monopoles are topological charges, they won't be able to annihilate in an environment with only one type of charges. 
Actually, even if thermal monopoles appear away from the central region, they will be created in pairs and thus be unable to affect the local charge imbalance.

This picture is confirmed in Figure \ref{fig5} where we simulate an initial heat pulse by introducing 10 pairs of monopoles at random positions in the central region, followed by a quench to low temperature (see appendix \ref{appMC} for details). We then use the same dynamics as before: no backtracking, charge creation is forbidden and domain walls are immobile. The goal is to measure how many monopoles persist at long times. This is what Figure \ref{fig5}(a) shows, after 400\,000 hopping attempts, averaged over 10\,000 independent runs. Depending on the density of domain walls $\rho_{dw}$, the 20 initial charges have a probability of around 40-50\% to persist, and more than 95\% of the simulations preserved at least 16 charges. These numbers should be taken only as qualitative indications though. We can expect them to vary noticeably depending on the precise parameters of the simulations e.g. with or without backtracking in the dynamics. But they allow us to extract a generic trend, namely that many monopoles persist at long times, and that increasing the density of domain walls tends to hinder annihilation.

This generic trend can be rationalised by comparing the two characteristic time scales of the monopole dynamics, namely the time $\tau_c$ for two opposite charges to meet, and the time $\tau_d$ to exit the central domain. The thickness of the central region is $\frac{L}{p+1}$, and for a given monopole, there are initially $n_c=10$ monopoles of opposite charge. The density of opposite charges in the central domain is $n_c\,(p+1)/L^3$, which gives a mean free path of $\ell_c=\frac{L}{[n_c (p+1)]^{1/3}}$. Since the number of monopoles in the central region can only decrease, this is a minimum value. In parallel, the shortest distance between a charge and the domain wall through which it can exit the central domain is, on average, $\ell_d=\frac{L}{2(p+1)}$. The ratio between the two time scales,
\begin{eqnarray}
\frac{\tau_c}{\tau_d}= \frac{\alpha_c}{\alpha_d}\sqrt{\frac{\ell_c}{\ell_d}}\propto \left(\frac{p+1}{\sqrt{n_c}}\right)^{1/3},
\label{eqtauratio}
\end{eqnarray}
tells us that for fixed system size $L$ and initial number of charges $n_c$, it becomes increasingly more likely to reach the domain wall before annihilation as $\rho_{dw}=\frac{p}{L}$ increases~\cite{footnote_prefactors}.

Figure \ref{fig5}(b) specifies the time evolution of this ensemble of carriers. At any given time, the distribution of positive (resp. negative) carriers spreads over the upper (resp. lower) domains, until they eventually accumulate on the upper (resp. lower) boundary of the system. In particular, we see that simulations after time $t\sim 100\,000$ hoppings are close to the asymptotic limit in time, which vindicates the results of Figure \ref{fig5}(a) that were taken after 400\,000 hoppings.

Our simulations and theory thus present the set-up of Figure \ref{fig4}(a) as a design able to accumulate magnetic charges of opposite signs on separate regions that are in practice insulated from each other; in other words a capacitor for magnetic monopoles in fragmented spin ice. Practically, either a constant heating or repeated heat pulses shall automatically increase the number of monopoles stored at the extremities. As a caveat, as the number of carriers increase, so will the number of paths for leakage. It means there can be a small but finite probability to go back to the previous region. More and more carriers are expected to populate regions just before the extremities (e.g. positive charges in regions $i=p-1, p-2 ...$), potentially introducing a density gradient along $z$. But as long as the density of carriers remains small compared to the domain-wall cross section, there is always a strong asymmetry for positive charges to go up rather than down (and vice-versa for negative ones). And with enough domain walls separating the two extremities, phase separation is strictly preserved.

\subsection{From a capacitor into a battery}
\label{sec:batt}

\begin{figure}
\centering
\begin{subfigure}[b]{0.58\columnwidth}
	\includegraphics[width=\columnwidth]{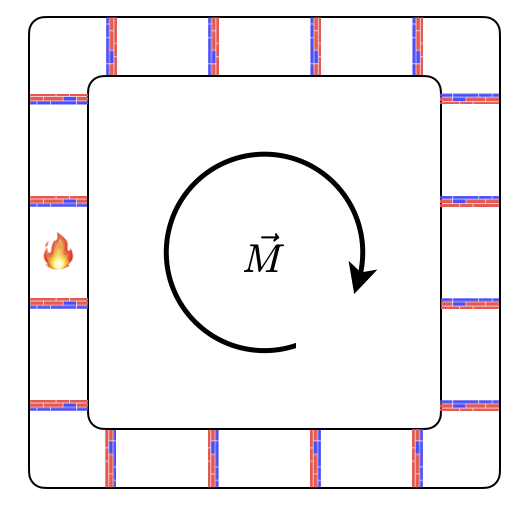}
	\caption{}
\end{subfigure}
\begin{subfigure}[b]{0.41\columnwidth}
	\includegraphics[width=\columnwidth]{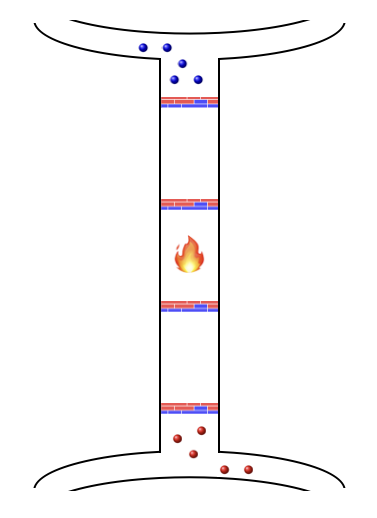}
	\caption{}
\end{subfigure}
\caption{
\textbf{Schematic illustrations of battery design:}
(a) In one circuit with domain walls. Positive and magnetic charges follow opposite directions thanks to the successive diodes, respectively clockwise and anti-clockwise. They eventually annihilate each other since they move on the same circuit. A constant heating (represented by a flame) is necessary to keep a flow of monopoles. Since the current flow always goes in the same direction, the leakage current continuously grows until it compensates the proper current and magnetricity stops; this is Lenz's law for magnetricity. The heating energy is transformed into a transient current of magnetic charges giving rise to a permanent macroscopic toroidal magnetic moment.
(b) In two external circuits without domain walls, connected to the reservoirs at the top and bottom. Since domain walls are only in the capacitor, but absent from the external circuits, an AC monopole current is now possible via an oscillating magnetic field (or possibly a time oscillating temperature gradient). The density of same-sign charges can be controled during the initial capacitor protocol, and is rigorously preserved in both circuits since monopole annihilation is impossible.
}
\label{fig6}
\end{figure}

With a capacitor at hand, a tempting question is how to use it as a battery ? In the circular setup of Figure \ref{fig6}(a), the heating on the left creates thermal monopole excitations in pairs, whose directions are then imposed by the domain walls: positive (resp. negative) carriers propagate clockwise (resp. counter-clockwise). Positive and negative charges shall eventually meet and annihilate in pairs, but a constant heating on the left will keep a continuous flow of carriers. The setup of Figure \ref{fig6}(a) thus transforms heat energy into magnetricity.

However, as mentioned in the introduction, a constant DC current of magnetic monopoles is not possible~\cite{jaubert09a,spinicebook}. The dynamics of a monopole comes with an attached Dirac string that flips spins along the way. Once all possible Dirac strings have been flipped, magnetisation has become saturated and there is no more monopole dynamics. After enough time, the setup of Figure \ref{fig6}(a) thus transforms heat energy into a toroidal magnetic moment whose direction is imposed by the $\mathbb{Z}_2$ broken symmetry of the domain walls.

Phenomenologically, the mechanism is an analogue of Lenz's law for magnetricity. Monopole dynamics creates a clockwise magnetisation, responsible for a counter-clockwise demagnetising field strength $\mathbf{H}$~\cite{jackson99a} which opposes the magnetic-charge dynamics that induced the magnetisation. Microscopically, the vanishing current comes from the leakage paths. Since carriers of a given sign always move in the same direction, the number of leakage paths necessarily increases with time until most of the domain-wall cross section has been crossed. The leakage current eventually becomes as important as the proper current and the monopole transport stops.


In a nutshell, since domain walls act as diodes, they can only induce a DC magnetic current in one given direction. And after a transient time, a DC magnetic current always stops. Hence, our theoretical proof of concept here is that domain walls can be used (i) as diodes, (ii) as capacitors and (iii) to create a battery of magnetic carriers, i.e. a reservoir for same-sign monopoles here. But another mechanism is necessary to induce a permanent magnetic current. This mechanism is an AC magnetic field, as well known in traditional spin ice~\cite{bramwell09a,jaubert09a,giblin11a,mengotti11a,kapaklis14a,farhan19a}. If each reservoir is independently connected to an external circuit made of a unique domain (either $\mathcal{D}$ or $\mathcal{\bar D}$), then an AC magnetic field will induce an AC current without the risk of the Lenz's effect since there are no domain walls in the external circuit [Figure \ref{fig6}(b)]. By working at low temperature where thermal excitations are prevented, our approach offers the possibility to control the density of monopoles in each circuit when charging the ``battery'' via the protocol of section \ref{sec:protoc}. And in absence of opposite-sign charges, there is no thermal noise due to creation/annihilation of monopoles.\\

As an aside, the present design may look like the magnetic analogue of a Brownian ratchet~\cite{Feynmanbook1}, with the apparent ability to produce work (here a battery for magnetic carriers) out of thermal noise. This is of course impossible. If the entire system were at the same temperature, then the appearance of monopole excitations means that excitations at the domain wall should also be possible. The domain wall, which plays the role of a Maxwell's demon in this thought experiment, would not be immobile anymore and its fluctuations would hinder the diode effect (see section \ref{sec:fluc}). This is why, in order to charge the capacitor, we need a heat source in the central region. This heating provides a thermal energy which is converted into magnetic energy as a battery of magnetic carriers -- or more prosaically, as a finite magnetic moment.

\section{Discussion}
\label{sec:disc}

\subsection{Experiments}

In light of our theoretical proof of concept, the next question is about how to realise these devices in experiments. Fragmentation has appeared in a variety of materials and experimental setups~\cite{lefrancois17a,Cathelin20a,museur26a,Canals16a,Sendetskyi16a,Yue22a,saccone23a,rougemaille19a,Petit16a,benton16b,Xu20a,paddison16a,dun20a,lhotel20a} and we shall focus on the two main options, namely pyrochlore oxides and artificial networks.

\subsubsection{Rare-earth iridium pyrochlore oxides}

Dy$_2$Ir$_2$O$_7$ and Ho$_2$Ir$_2$O$_7$ are constituted of two interpenetrated pyrochlore lattices made of rare-earth ions on one hand (Dy or Ho), and transition-metal ions on the other hand (Ir). The crystal field on Dy$^{3+}$ and Ho$^{3+}$ in pyrochlore oxides is responsible for a strong Ising anisotropy~\cite{gardner10a,lefrancois17a,Cathelin20a}. Alone on the pyrochlore lattice, Dy/Ho spins support traditional spin-ice physics~\cite{Harris97a,spinicebook}. But once they are coupled to Ir spins, fragmented spin ice takes place~\cite{lefrancois17a}. For the most part, this is due to the large exchange coupling between 5d ions ($\sim 100$K), two orders of magnitude larger than between 4f ions ($J_{eff} \sim 1$K). As a consequence, Ir magnetic moments order at relatively high temperatures, between $120-140$ K depending on the material~\cite{bansal02a,matsuhira11b}. This long-range order induces a staggered local field $h_{loc}$ on the rare-earth Ising moments which competes with the traditional spin-ice ground state induced by $J_{eff}$. For $h_{loc}/J_{eff}<2$, the ground state of Dy/Ho spins would remain traditional spin ice. For $h_{loc}/J_{eff}>6$, the Dy/Ho ground state would become AIAO order. But for the window $2<h_{loc}/J_{eff}<6$, the Dy/Ho lattice supports fragmented spin ice~\cite{lefrancois17a}. This is what happens in Dy$_2$Ir$_2$O$_7$ and Ho$_2$Ir$_2$O$_7$ where $h_{loc}/J_{eff}\approx 4.5$. Let us now discuss if these two materials could fit the design for diodes and capacitors of magnetricity.

Among the conditions assumed in our work, we have considered that the chemical potential of magnetic charges was the same for 2in-2out and 4in or 4out excited monopole states. That's precisely the case at $h_{loc}/J_{eff}= 4$~\cite{lefrancois17a}, which is fortunately reasonably close to the estimate of $h_{loc}/J_{eff}\approx 4.5$ in Dy$_2$Ir$_2$O$_7$ and Ho$_2$Ir$_2$O$_7$. Furthermore, during its dynamics, a monopole (say a positive one) becomes alternatively ... 2in-2out / 4in / 2in-2out / 4in ... Hence any energy cost due to this difference of chemical potential is recovered every two monopole hoppings~\cite{jaubert15c,lefrancois17a}. This might slow down the dynamics, introducing a different time scale every other hopping, but it is not expected to affect the property of domain walls since their asymmetric filtering lies in their topology. Similarly, magnetic dipolar interactions between rare-earth ions famously renormalise into a magnetic Coulomb potential between monopole excitations~\cite{castelnovo08a,jaubert15c}. But since the Coulomb potential is not confining~\cite{castelnovo08a}, this is also not expected to qualitatively change our results.

Since magnetic fragmentation is stabilised by the local field coming from the iridium ions, it means that fragmented domains of the Dy/Ho spins are, to a first approximation, attached to AIAO iridium domains. In other words, the fragmented domain $\mathcal{D}$ (resp. $\mathcal{\bar D}$) is favoured by the AIAO domain $\mathcal{D}_{\rm Ir}$ (resp. $\mathcal{\bar D}_{\rm Ir}$) on the iridium lattice. It has been confirmed in Ho$_2$Ir$_2$O$_7$~\cite{Pearce22a} where a strong [111] magnetic field selects a unique state belonging to one of the two fragmented domains of Ho ions (say $\mathcal{D}$); in response, as domain $\mathcal{D}$ becomes macroscopic, it puts energetic pressure for the growth of domain $\mathcal{D}_{\rm Ir}$ on the iridium lattice. Once the system is cooled down to sub-Kelvin temperature $T_s$, the magnetic field can be suppressed. With an energy scale of more than 100K, domain $\mathcal{D}_{\rm Ir}$ should persist (and being antiferromagnetic, there is no demagnetisation field to induce domains), while the fragmented state on the Ho lattice can relax via monopole dynamics, whose density can be controlled with $T_s$. Then cooling further down to low enough $T_f<T_s$, (most of) the monopole excitations are suppressed and we obtain a seed configuration into domain $\mathcal{D}$.

The difference of exchange energy scale between Ir spins on one hand and Dy/Ho spins on the other hand plays an important, fortunate, role. Firstly, at sub-Kelvin temperatures, monopole hopping on Dy/Ho sites won't affect the underlying AIAO iridium domains. Secondly, a temperature increase as small as 100 mK would easily create monopole excitations~\cite{jaubert11c} on the Dy/Ho pyrochlore lattice ($J_{eff}\sim 1$ K~\cite{lefrancois17a,Cathelin20a}, but would be essentially irrelevant to Ir ions which order at $120-140$ K. The local heating necessary in the design of the capacitor (see section \ref{sec:capac}) and battery (see section \ref{sec:batt}) would thus not affect the underlying $\mathcal{D}_{\rm Ir}$ domain of Ir spins, whose local field $h_{loc}$ would bring back domain $\mathcal{D}$ on the Dy/Ho lattice once the heating is stopped. Finally, establishing a domain wall between $\mathcal{D}_{\rm Ir}$ and $\mathcal{\bar D}_{\rm Ir}$ iridium domains would induce a domain wall at the same position between fragmented Dy/Ho domains $\mathcal{D}$ and $\mathcal{\bar D}$. The condition to respect would then be to work at temperatures low enough, $T \ll h_{loc}\approx 5-6$ K in Dy$_2$Ir$_2$O$_7$ and Ho$_2$Ir$_2$O$_7$, to prevent Dy/Ho domain wall fluctuations. The advantage of the designs presented in this work is that once monopoles have been thermally created and have migrated beyond at least one domain wall, the temperature can be cooled down as low as necessary (as long as hopping dynamics is not frozen); the diode effect will prevent opposite charges to meet again and thus prevent their annihilation. Thanks to their topological nature, monopoles will persist even at temperatures much lower than their chemical potential.

An advantage of AIAO order is that even though it is antiferromagnetic, it is made of ferromagnetic layers, which are successively pointing into opposite directions. Its response to a magnetic field is non-trivial and specific AIAO domains have been selected via different experimental probes in a variety of pyrochlore oxides~\cite{Tardif15a,Fujita15a,Opherden17a,Pearce22a}. Manipulating an AIAO domain wall is thus possible. The last ingredient is to make this AIAO iridium domain wall flat, within one tetrahedron layer. This is the main technological challenge to overcome in order to make our theoretical proof of concepts into solid-state experiments. A possible approach might be in thin films, that are possible in pyrochlore oxides~\cite{bovo14a,Leusink14a,Fujita15a,Fujita16a,Bovo17a}. A similar idea has been proposed to inject monopoles from boundary effects in traditional spin ice~\cite{Miao20a,Timsina25a}. Here, one could take advantage of the successive ferromagnetic layers of AIAO order. Once a given domain $\mathcal{D}_{\rm Ir}$ is established in the entire system, surface effects would impose boundary conditions locally favouring the opposite $\mathcal{\bar D}_{\rm Ir}$ domain. The advantage is that once boundary conditions impose the proximity of two different domains, a low temperature would naturally favour a flat domain wall -- especially in a thin film whose penetration depth can be controlled -- as its flatness minimises the interface energy pressure.

As a summary, Dy$_2$Ir$_2$O$_7$ and Ho$_2$Ir$_2$O$_7$ present several fortuitous features making them promising materials for our proposal. The flatness of the domain wall remains an open question, but directions of investigations are possible in thin films and heterostructures

\subsubsection{Artificial networks}

Stepping away from solid-state physics, magnetic fragmentation has also been sought after in artificial spin ice~\cite{wang06a,spinicebook}. Artificial lattices are realised by nano-lithography, typically as an array of micrometer-size islands of permalloy, deposited on a non-magnetic substrate of silicium. While initially inspired by spin ice, the field has evolved far beyond into metamaterials, reprogrammable magnonic crystals and neuromorphic computing~\cite{skjaervo19a}. Its main advantage is the flexibility of the design, recently extended to three dimensions by metallic deposition on 3D polymer nanoprinting~\cite{may19a,May21a,saccone23a,berchialla24a}, and room-temperature experiments. The diversity of artificial-lattice geometries is immense~\cite{skjaervo19a} and essentially only limited by the imagination of their author.

In this context, it is maybe not surprising if fragmentation was first observed in artificial spin ice~\cite{Canals16a,Sendetskyi16a,Yue22a,rougemaille19a,saccone23a}. However, most efforts have been devoted to the kagome geometry, where fragmentation naturally appears due to dipolar interactions~\cite{moller09a,chern11a,rougemaille19a}. Unfortunately, domain walls on fragmented kagome ice do not share the property of their pyrochlore counterparts \footnote{The kagome lattice is made of corner-sharing tetrahedra. When creating a domain wall across a triangle, flipping two spins of a 2in-1out or 2out-1in state might induce a 3out or 3in spin configurations, thus creating an excitation.} and thus cannot be directly applied to our proof of concept. But all of our arguments do apply to domain walls on 2D fragmented square ice. In particular the microscopic mechanisms discussed in section \ref{sec:mech} extends fairly straightforwardly to two dimensions. The only difference would be on the dynamics of monopoles, as random walks behave differently in 2D and 3D. In two dimensions, monopoles would a priori more quickly reach the opposite domain wall in the capacitor design of Figure \ref{fig4}(a), thus increasing their effective velocity. But at the same time, the probability of returning to origin is higher in two dimensions, which is expected to increase the probability of the leakage current. But the asymmetric filtering would be the same, protected by the topology of the lattice. In addition, fragmentation has recently been predicted numerically and observed experimentally in 3D artificial spin ice~\cite{saccone23a}, in a geometry equivalent to pyrochlore. 

In any case, artificial spin ice presents several advantages. Room-temperature is an important one, as well as the increase of the length scale by a factor of $\sim 10^3$, from a few nanometers to a micrometer. As a consequence, artificial lattices offer more tools for manipulation of the degrees of freedom, and even a direct observation of the charge excitations by magnetic force microscopy, Lorentz microscopy, photo-emission electron microscopy ... And if needed, the network can be redesigned, finely tuned for example to impose a local field where domains need to be flipped in order to create domain walls. The scenarii are numerous and, in light of the past 20 years of research in the field, we are reasonably optimistic that our proposal of a domain wall between fragmented-spin-ice phases can be realised in artificial lattices, either in 2D or in 3D.

As an interesting alternative in the future, we would finally like to mention the realisation of ``qubit spin ice'' in a quantum annealing system~\cite{king21a}. While that work focused on reproducing the Coulomb phase of traditional spin ice, the flexibility of quantum annealers should enable the stability of a fragmented phase, and where the presence of a domain can be encoded directly.

\begin{figure}[b]
\centering\includegraphics[width=\columnwidth]{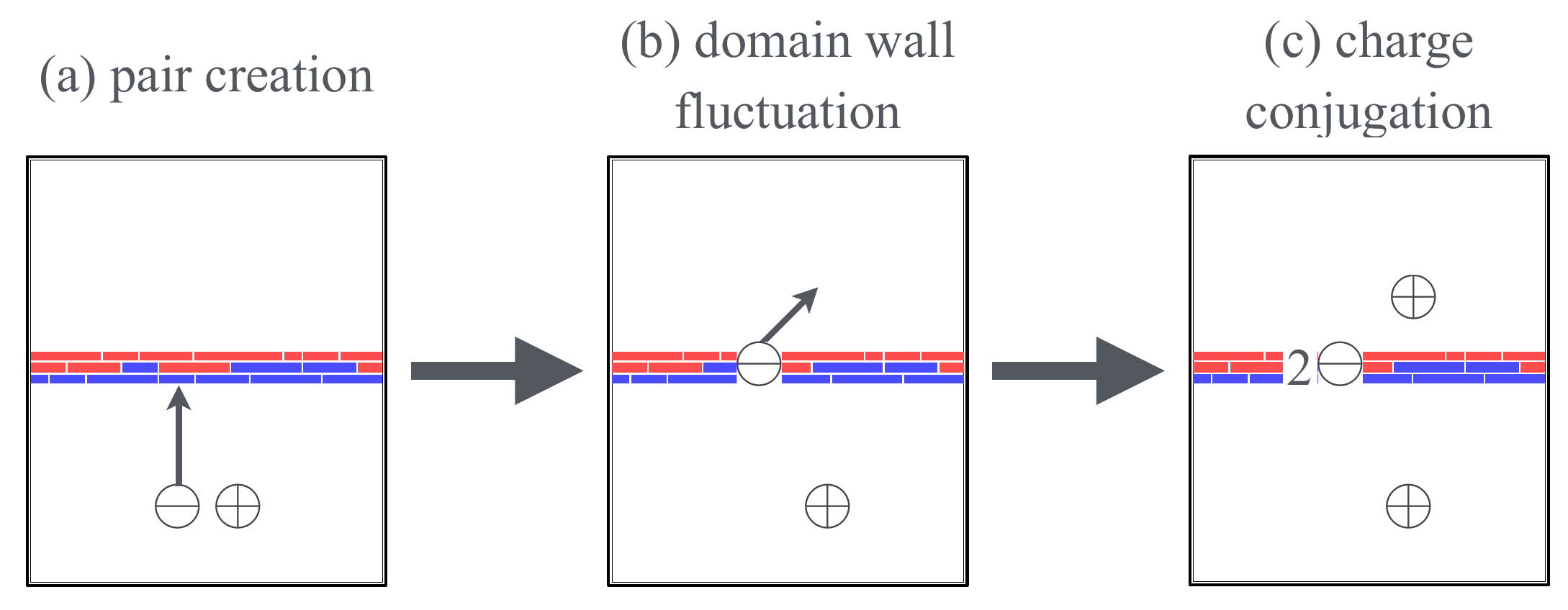}
\caption{
\textbf{Domain wall fluctuations}
(a) After a pair of charge excitations is created in domain $\mathcal{D}$, the negative charge reaches the domain wall. (b) Here we allow the domain wall to fluctuate. This fluctuation enables the negative charge to cross. (c) The fluctuation is equivalent to creating a double charge inside the wall. In order to preserve charge neutrality, the hopping charge has to change sign. For more details, see section \ref{sec:fluc}.
}
\label{fig7}
\end{figure}

\subsection{Domain wall fluctuations and charge conjugation}
\label{sec:fluc}

Before concluding, there is one last point we would like to discuss. We have so far always imposed immobile domain walls. Spins can of course flip at the interface -- otherwise charges could not go through -- but the boundary between domains $\mathcal{D}$ and $\mathcal{\bar D}$ remained fixed. We've seen in section \ref{sec:mech} that it was a necessary condition for the diode effect due to domain walls. But what happens if domain walls \textit{can} fluctuate ? We shall first describe the microscopic mechanism of this fluctuation, before presenting the physics intuition.

Imagine a negative charge entering a domain wall through a 3in-1out tetrahedron from domain $\mathcal{D}$, as illustrated in section \ref{sec:mech} and Figure \ref{fig3}(d,e). Instead of being reflecting by exiting through an outward spin as was previously the case, let us allow our negative charge to flip an inward spin. The charge is now able to cross the wall and penetrates domain $\mathcal{\bar D}$, leaving behind a 3out-1in tetrahedron. This tetrahedron has shifted domains (from $\mathcal{D}$ to $\mathcal{\bar D}$), representing a local fluctuation of the domain wall. The subtle consequence is that the carrier has now exited a tetrahedron via an \textit{inward} spin, which is not the behaviour of a negative charge, but rather of a positive one. Domain wall fluctuations is equivalent to charge conjugation on the magnetic carrier !

Intuitively, the wall fluctuation can be seen either as a tetrahedron shifting from domain $\mathcal{\bar D}$ to domain $\mathcal{D}$; or as a local double-charge excitation in domain $\mathcal{\bar D}$ (from 3in-1out state to 3out-1in). In order to conserve the overall charge neutrality of the system, this double charge must be accompanied by a change of sign of the hopping carrier. In other words, the dynamical negative charge becomes positive after crossing the interface, leaving a double negative charge inside the wall, as illustrated in Figure \ref{fig7}. But since this double charge is exactly at the boundary, it can just as well be seen as a ground state of the other domain $\mathcal{D}$; hence, a domain wall fluctuation.  From this point of view, we are left with two positive charges in the system, and one domain-wall fluctuation. Note that the same reasoning applies to positive charges.

Since this mechanism is a direct consequence of the total charge conservation in the system, it is independent of the shape of the interface. Domain wall fluctuations break the diode effect, but they do so at the cost of charge conjugation on the carrier traversing the wall; a property that will affect the magnetic dynamical properties of fragmented spin-ice materials~\cite{lefrancois17a,Cathelin20a,Pearce22a,museur26a}.

\subsection{The domain wall is a frontier between two worlds}
\label{sec:mech}

As mentioned in the introduction, the domain wall can be seen as separating two time-reversal symmetric worlds supporting their own form of electromagnetism, namely the Coulomb phase. From that point of view, the immobile domain wall only allows half of the matter to cross in a given direction (i.e. the diode effect): positive on one side and negative on the other side. Since charge excitations are topologically stable and can only annihilate in pairs of opposite signs, once they are split on either side of the wall, they will persist over a long period of time.

Pushing this emergent analogy further, the domain wall effectively results in a separation between what can be seen as classical analogues of matter (say the positive charges) and anti-matter (the negative ones). The analogy would be somewhat irrelevant in absence of a domain wall, since in spin ice, or in a single domain of fragmented spin ice, the underlying Coulomb phase naturally describes excitations as a gas of positive and negative charges. But since the domain wall separates monopoles of opposite charge, matter and anti-matter are now independently (and topologically) stable in their own respective world (i.e. domain). They can evolve and be manipulated independently from each other. This is of course nothing more than a classical analogy emerging from our spin model, but it represents an intuitively simple description of the physics in presence of an immobile domain wall in fragmented spin ice.\\

Pursuing the analogy a bit more, we see that a soft, fluctuating,  boundary between the two worlds (i.e. the two domains), does not spontaneously split matter from anti-matter anymore (i.e. positive from negative charges). However, let us assume for the sake of argument that our system was prepared with all of matter on one side, and all of anti-matter on the other side. For example this could be done by keeping the appropriate domain wall immobile for long enough time, until positive and negative charges are separated. Then matter separation would persist even if the interface becomes soft. Indeed, matter separation is automatic if the domain wall remains locally immobile during crossing; this is the diode effect. And if the interface fluctuates, then positive charges become negative when crossing the wall, and vice-versa. It means that we can expect the amount of matter to vary in each domain over time, but, aside from the influence of leakage paths, their sign (i.e. their matter or anti-matter nature) will stay the same on each side.

\section{Conclusion}

We have studied the influence of a domain wall on the dynamics of magnetic monopole excitations in fragmented spin ice. By breaking time reversal symmetry, passing through the wall confers a preferred direction to the monopole dynamics, attached to its topological (and magnetic) charge. A given immobile domain wall will systematically reflect a specific charge but will let through its opposite counterpart. In the context of a current of magnetic monopoles (a.k.a.``magnetricity''~\cite{bramwell09a}), this domain wall can serve as a diode. Adding several domain walls one after the other, in the right order, induces a clean charge separation equivalent to a capacitor. This capacitor offers two distinct baths of monopoles: positive charges on one end, and negative ones on the other end. Being topological in nature, since positive and negative charges are separated, they cannot annihilate, but can be used as a battery of magnetic carriers. Our work thus offers a theoretical proof of concept for the design of complex devices in the transport of magnetic monopoles. Possible experimental realisations are discussed in iridium pyrochlore oxides and artificial lattices. A soft domain wall, where fluctuations are possible, does not preserve the diode effect. Instead, fluctuations induce a systematic charge conjugation when the carrier crosses the interface.

As an emergent gauge field theory, the domain wall represents a separation between two time-reversal symmetric worlds (i.e. domains $\mathcal{D}$ and $\mathcal{\bar D}$), each one described by its own simplified Maxwell's equations. From this point of view, the diode effect of an immobile domain wall separates an analogue of matter from anti-matter, since positive charges (matter) are topologically stable and cannot vanish without their negative counterparts (anti-matter).

Beyond the transport of magnetic monopoles, the generic physics of domain walls in disordered magnetic textures such as spin liquids remains largely uncharted, perhaps because the two concepts are a priori antinomic. A natural direction to take these ideas forward would be to investigate how they would operate in the context of quantum fragmented spin ice, where Maxwell's equations gain dynamics and dynamical axions are expected~\cite{Pace23a}. But fragmentation is not the only example of a spin liquid with broken symmetry. Famous examples include chiral~\cite{Kalmeyer87a,Bauer14a,He14a,Bieri15a,Essafi16a} or nematic spin liquids~\cite{Grover10a,Shannon06a,Thomson18a,Benton18a}. In such spin liquids, domain walls are not only tools to characterise the transition (often done by measuring its tension~\cite{Sorokin12a,Szasz20a}), but also come with their own exotic properties~\cite{wang22a}. If spin liquids can be seen as their own little worlds described by an emergent gauge field, domain walls offer a tangible and porous frontier between distinct gauge fields, where quasi-particle excitations are the perfect probe. If the broken symmetry is $\mathbb{Z}_2$, then the two distinct domains are, to some extent, opposite from each other. But magnetic order with higher degeneracies ($\mathbb{Z}_3, \mathbb{Z}_6$, or even U(1)) would lead to more complex broken symmetry, and domain walls would come with multiple flavours that potentially opens an exciting horizon.\\\\

\textit{Acknowledgements:} 
We acknowledge the computational resources of the Param RUDRA and Chandra HPC clusters.
A.R. and S.P. thank Kedar Damle and G J Sreejith for discussions.
L.D.C.J.\ thanks Edgar Bonet for insightful questions, and acknowledges support from the French National Agency for Research (ANR-23-CE30-0038-01 and ANR-25-CE30-5029-04) and from Idex Bordeaux (Research Program GPR Light).
SP acknowledges support from ANRF-DST (formerly SERB), Govt. of India via Grant No. MTR/2022/000386.
S.P. also acknowledges the Indo-Japan LOTUS science exchange award for partial support during the final stages of this project.

\appendix
\section{Details of the simulations}
\label{appMC}

In order to build our analysis on an unbiased method, we used the numerical protocol that has been validated in Refs.~\cite{castelnovo11a,jaubert11c,jaubert15c} when confirming the Debye-H\"uckel theory and Coulomb potential between magnetic monopoles in spin ice. Here we adapted this protocol to the presence of domain walls in the fragmented phase.

At first, one needs to choose a fragmented spin configuration. A trivial one consists of all spins pointing towards the [111] direction~\cite{matsuhira02a}. This state is, however, fully ordered. In order to obtain a random fragmented state, $10^4$ random loop updates are then performed~\cite{brooks14a}. Loop updates are necessary to respect the zero-divergence condition, since local spin flips would create charge excitations. 

In Figure \ref{fig3}, we performed simulations on a pyrochlore lattice made of more than one million spins ($N=4\,L^3$ spins with $L=72$). The system is split into two domains $\mathcal{D}$ and $\mathcal{\bar D}$ by flipping all spins above the domain wall at $z>L/2=36$. First, one creates a single pair of monopoles by flipping one spin in the centre of the lower domain~\cite{jaubert15c}. This is our seed configuration. Then we choose which charge, either the positive or the negative one, will be moved. The other one stays immobile until annihilation. This way we can study the dynamics of one type of charge, without perturbation of the other one on the domain wall, as if the moving charge was essentially alone for most of its dynamics. At each step, this charge is randomly moved to a neighbouring tetrahedron by flipping a spin. Here we assume that all moves are allowed as long as:
\begin{enumerate}[label=(\roman*)]
\item they do not create another monopole excitation. The number of monopole is thus always two: one fixed and one moving.  This corresponds to a very-low-temperature dynamics
\item they do not modify the shape of the domain wall. Any given tetrahedron always remains in the same domain, either $\mathcal{D}$ or $\mathcal{\bar D}$, throughout the monopole dynamics, as defined by its seed configuration.
\item they do not backtrack. This choice is not strictly necessary, but it improves the efficiency of the algorithm. It means that if the monopole hops from tetrahedron A to tetrahedron B, then it cannot immediately hop back to tetrahedron A. The advantage is to prevent the possibility for the two charges to annihilate immediately at the first step. The shortest path for annihilation is a loop of size six.
\end{enumerate}
At each step, we update the probability distribution as a function of the layer $z$ where the moving monopole is. The iteration stops when the two monopoles meet again and annihilate. This diffusion process flips a closed loop of spins, from creation to annihilation of the monopole pair. A loop can be as small as 6 steps, but the fact that a random walk is transient in three dimensions ensures that most loops are of the order $10^4-10^6$ steps~\cite{jaubert11a}. Our results are then averaged over $10^3$ monopole diffusion (i.e. loops) for each of the $200$ uncorrelated seed configurations.\\

In Figure \ref{fig4}, system size is $L=256$. The data are averaged over 20,000 independent runs.\\

In Figure \ref{fig5}, system size is $L=256$. At first, ten spins are randomly flipped, creating ten pairs of charges in the central domain. The initial state at time $t=0$ is thus filled with positive and negative charges sitting next to each other, As a consequence, an artificially high number of pair annihilation should be relatively frequent at very short time. However, the absence of backtracking in our simulations prevents such immediate annihilation. Monopoles are forced to diffuse, at least around an hexagonal plaquette, before annihilation. In practice, the absence of backtracking provides a randomly diluted gas of charges after just a few steps, which is a realistic seed configuration. At each step, a monopole is chosen randomly and moved randomly following the same dynamics as before. The novel aspect is that if this monopole hits another monopole of opposite charge, then they annihilate. This corresponds to a very-low-temperature dynamics, where excitations can annihilate but not be created. In panel (a), data are averaged over 10,000 independent runs, and each run lasts for $400\,000$ hopping attempts. In panel (b), data are averaged over 20,000 independent runs, and each run lasts for $100\,000$  hopping attempts.\\

Note that periodic boundary conditions in simulations impose a second domain wall at the boundary along the $z-$direction. To avoid numerical artefacts with monopole carriers wrapping around the system artificially, this boundary domain wall at $z=L$ will always be defined as reflective for both charges; in other words, no carriers shall cross this domain and hop from $z=L$ to $z=0$. For the purpose of this paper, this is essentially equivalent to open boundaries along the $z-$direction.

\bibliography{biblio2}
\end{document}